\documentclass[twocolumn]{aa}
\usepackage{graphicx}
\usepackage[intlimits,sumlimits]{amsmath}
\usepackage{deluxetable}
\usepackage{times,epsfig} 
\usepackage{natbib}
\usepackage{amssymb}
\usepackage{mathrsfs}  
\bibpunct{(}{)}{;}{a}{}{,} 
\usepackage{caption}

\usepackage{amsmath}
\usepackage[english]{babel}
\usepackage{longtable}
\usepackage{url}
\usepackage{hyperref}

\newcommand{\beqa}{\begin{eqnarray}} 
\newcommand{\eeqa}{\end{eqnarray}}

\newcommand{\bsub}{\begin{subequations}}
\newcommand{\esub}{\end{subequations}}
\newcommand{\beal}{\begin{align}}
\newcommand{\ealn}{\end{align}}
\newcommand{\Nif}{$\rm ^{56}Ni$} 
\newcommand{\Cif}{$\rm ^{56}Co$}

\begin{document}

\title{The normal Type Ia SN~2003hv out to very late phases
\thanks{Based on observations collected at the European Southern
    Observatory, Paranal, Chile (ESO Programmes 073.D-0294(AB),
    074.D-0259(BC) and 075.D-0116(AB)), Cerro Tololo Inter-American
    Observatory, Lick Observatory, Las Campanas Observatory, Siding
    Spring Observatory, and the {\em Hubble Space Telescope}.}}

\author{
G.~Leloudas\inst{1} 
\and M.~D.~Stritzinger\inst{2,1} 
\and J.~Sollerman\inst{1,3} 
\and C.~R.~Burns\inst{4}
\and C.~Kozma\inst{3} 
\and K.~Krisciunas\inst{5}
\and J.~R.~Maund\inst{1}
\and P.~Milne\inst{6}
\and A.~V.~Filippenko\inst{7} 
\and C.~Fransson\inst{3} 
\and M.~Ganeshalingam\inst{7}
\and M.~Hamuy\inst{8} 
\and W.~Li\inst{7}
\and M.~M.~Phillips\inst{2}
\and B.~P.~Schmidt\inst{9}
\and J.~Skottfelt\inst{10} 
\and S.~Taubenberger\inst{11}
\and L.~Boldt\inst{2}
\and J.~P.~U.~Fynbo\inst{1}
\and L.~Gonzalez\inst{8}
\and M.~Salvo\inst{9}
\and J.~Thomas-Osip\inst{2}
}

\institute{Dark Cosmology Centre, Niels Bohr Institute, University of
  Copenhagen, Juliane Maries Vej 30, 2100 Copenhagen \O,
  Denmark;\\ \email{giorgos@dark-cosmology.dk}
\and Las Campanas Observatory, Carnegie Observatories, Casilla 601, La
Serena, Chile
\and The Oskar Klein Centre, Department of Astronomy, Stockholm University, 
AlbaNova, 10691 Stockholm, Sweden
\and Observatories of the Carnegie Institution of Washington, 813 Santa Barbara 
St., Pasadena, CA 91101, USA 
\and Department of Physics, Texas A\&M University, College Station, Texas 77843, 
USA 
\and Department of Astronomy and Steward Observatory, University of Arizona, 
Tucson, AZ 85721, USA
\and Department of Astronomy, University of California, Berkeley, CA
94720-3411, USA
\and Universidad de Chile, Departamento de Astronom\'{\i}a, Casilla 36-D, 
Santiago, Chile 
\and Research School of Astronomy and Astrophysics, Australian
National University, Cotter Road, Weston Creek, PO 2611, Australia
\and Niels Bohr Institute, University of Copenhagen, Blegdamsvej 17,
2100 Copenhagen \O, Denmark
\and Max-Planck-Institut f\"ur Astrophysik, Karl-Schwarzschild-Str.~1,
D-85741 Garching bei M\"unchen, Germany 
}

\offprints{G. Leloudas}
\date{Received 22 April 2009 / Accepted 7 July 2009}

\abstract{}
{
We study a thermonuclear supernova (SN), emphasizing very late phases.
}
{{
An extensive dataset for SN~2003hv that covers the flux evolution from
maximum light to day $+$786 is presented.  This includes 82 epochs of
optical imaging,  24 epochs of near-infrared imaging, and 10 epochs of
optical spectroscopy.  These data are combined with published
nebular-phase infrared spectra, and the observations are compared to
model light curves and synthetic nebular spectra.
}}
{{
SN~2003hv is a normal Type Ia supernova (SN~Ia) with photometric and
spectroscopic properties consistent with its rarely observed $B$-band
decline-rate parameter, $\Delta m_{15}(B) = 1.61 \pm 0.02$.  The
blueshift of the most isolated [\ion{Fe}{ii}] lines in the
nebular-phase optical spectrum appears consistent with those observed
in the infrared at similar epochs.  At late times there is a prevalent
color evolution from the optical toward the near-infrared bands.  We
present the latest-ever detection of a SN~Ia in the near-infrared in
{\it Hubble Space Telescope} images. The study of the
ultraviolet/optical/infrared (UVOIR) light curve reveals that a
substantial fraction of the flux is ``missing'' at late times.
Between 300--700 days past maximum brightness, the UVOIR light curve declines
linearly following the decay of radioactive \Cif, assuming full and
instantaneous positron trapping.  At 700 days we detect a possible
slowdown of the decline in optical bands, mainly in the $V$ band.
}}
{
The data are incompatible with a dramatic infrared catastrophe.
However, the idea that an infrared catastrophe occurred in the 
densest regions before 350 days can explain the missing flux from the
UVOIR wavelengths and the flat-topped profiles in the near-infrared.
We argue that such a scenario is possible if the ejecta are clumpy.
The observations suggest that positrons are most likely trapped in 
the ejecta.
}
{
}

\keywords{supernovae: general -- supernovae: individual: SN~2003hv}

\maketitle

\section{Introduction}

Studying the late-phase emission of Type Ia supernovae (hereafter
SNe~Ia) provides an excellent opportunity to elucidate the physical
nature of these thermonuclear explosions.  With late-time observations
it is in principle possible to constrain the nucleosynthetic yields
and the distribution of elements
\citep{2005A&A...437..983K,2006ApJ...652L.101M,2006A&A...460..793S,2007Sci...315..825M},
the magnetic field configuration of the ejecta and hence of the
progenitor
\citep{1980ApJ...237L..81C,1998ApJ...500..360R,1999ApJS..124..503M},
and potentially their contribution to the diffuse Galactic 511 keV
line \citep{1999ApJS..124..503M,2008NewAR..52..457P}.

Previous studies of the late-time emission of SNe~Ia have demonstrated
the increased importance of the near-infrared (NIR) emission. This
investigation builds upon earlier studies of SN~2000cx
\citep[][hereafter S04]{2004A&A...428..555S} and SN~2001el
\citep[][hereafter SS07]{2007A&A...470L...1S} but extends to even
later phases. We present observations of SN~2003hv that cover 786 days
past $B$-band maximum ($B_{\rm max}$).  One of the main motivations of
this study was to investigate whether an infrared catastrophe (IRC)
occurs in the ejecta at these very late phases.  The IRC is a thermal
instability, predicted by \cite{1980PhDT.........1A}, which shifts the
bulk of the emission from the optical and NIR regime to the mid- and
far-infrared fine-structure lines once the temperature in the ejecta
falls below a critical limit.  While the IRC has been proposed to
explain the line evolution of SN~1987A
\citep{1992MNRAS.255..671S,1998ApJ...496..946K}, it has never been
observed for a thermonuclear SN \citep[see, however,][on the peculiar
  SN~2006gz]{2009ApJ...690.1745M}. Models suggest that the IRC should
commence, depending on the structure and composition of the ejecta,
sometime between 500--700 days past maximum brightness and become
apparent with a dramatic drop in the optical and NIR luminosity. This
corresponds to shortly after the end of our previous multi-band
observations of other SNe~Ia (S04; SS07).

SN~2003hv was discovered by LOTOSS \citep{2003IAUC.8197....1B}  on
9.5 September 2003 (UT dates are used throughout this paper).  It was
located 17$\arcsec$ east and 57$\arcsec$ south of the nucleus of the
S0 galaxy NGC 1201.  \citet{2003IAUC.8198....2D} classified it as a
SN~Ia the day after discovery.  SN~2003hv reached an apparent $B$-band
magnitude of 12.45 mag, thus being the brightest supernova discovered
in 2003, and one of the brightest SNe~Ia observed over the past
decade.  NGC~1201 has a direct distance measurement that is based on
the surface brightness fluctuations method
\citep[SBF;][]{2001ApJ...546..681T}.  By applying a correction of
$-$0.16 mag \citep{2003ApJ...583..712J}, the distance modulus of
NGC~1201 is $\mu_{\rm SBF} = 31.37 \pm 0.30$ mag, corresponding to a
distance of 18.79 $\pm$ 2.60 Mpc.  The Galactic extinction in the
direction of NGC~1201 is low, $E(B-V)=0.016$ mag
\citep{1998ApJ...500..525S}, and we show below that there is no
evidence of any host-galaxy extinction.  As SN~2003hv suffered little
extinction, was located in the outskirts of its host, and was bright,
it made an excellent target for our late-time photometric ($UBVRIJHK$)
and spectroscopic (optical) observational campaign.

SN~2003hv has already attracted some attention in the
literature. \citet{2006ApJ...652L.101M} published a NIR spectrum taken
394 days past maximum brightness. They found that the [\ion{Fe}{ii}]
1.644 $\mu$m emission line exhibited a flat-topped profile and was
blueshifted by $\sim$2600 km~s$^{-1}$.  \citet{2007ApJ...661..995G}
showed a mid-infrared (MIR) spectrum (5.2$-$15.2 $\mu$m) of SN 2003hv
taken on day $+$358 with the InfraRed Spectrograph onboard the 
{\em Spitzer Space Telescope}.  From the observed [\ion{Co}{iii}]
emission they estimated that $\sim$0.5 M$_{\sun}$ of \Nif~was
synthesized in the explosion.  In this paper, we present a
comprehensive study of SN~2003hv that covers the flux evolution from
early to very late phases.  Some preliminary results of our study
  were presented by \cite{2009AIPC.1111..456L}. Here, we have used
  more optical and NIR early-time data, have re-analysed all data in a
  consistent manner, and have supplemented our late-time analysis with
  data from the {\em Hubble Space Telescope (HST)} and
  spectral-synthesis modeling; we have also included the spectra
  published by \citet{2006ApJ...652L.101M} and
  \citet{2007ApJ...661..995G}.

The organization of this paper is as follows.
Section~\ref{sec:Obs} presents the
observations that were collected using ground-based facilities
and {\it HST}, and Sect.~\ref{sec:red}
describes the corresponding data reduction. The results are described
in Sect.~\ref{sec:res}. Section~\ref{sec:disc} provides a discussion, 
and the main conclusions are summarized in Sect.~\ref{sec:conc}.

\section{Observations}
\label{sec:Obs}

The observations presented in this paper were conducted at optical 
and NIR wavelengths with a variety of facilities. 
A brief description of the observations is given below.
Observing logs for our data are found 
in Table~\ref{tab:spectralog} and in 
Tables~\ref{tab:VLTopticallog} -- \ref{tab:HSTlog} of the Appendix.

\begin{table*}[tbh]
\centering
\caption[]{Log of spectroscopy.}
\label{tab:spectralog}
\begin{tabular}{cccllcc}
\hline\hline
Date & MJD & Phase$^a$ & Telescope & Instrument  & Resolution &Exposure time \\ 
 (UT) &(days) &   (days)     &      &     & (\AA) &  (s)                    \\
\hline
2003 09 10    &  52892.35     &     1.2     &	 LCO  Baade	  &  IMACS    &    23.0      &  $2 \times 300$  \\
2003 09 15    &  52897.34     &     6.1     &	 LCO  Clay	  &  LDSS2    &   13.5  &   $1 \times 60$  \\ 
2003 09 18    &  52900.39     &     9.2     &	 LCO  duPont	  &  ModSpec    &     6.0  &   $2 \times 300$ \\ 
2003 09 26    &  52908.31     &    17.1     &	 LCO  duPont	  &  WFCCD    &     6.0  &    $2 \times 300$  \\ 
2003 10 30    &  52942.6      &    51.4     &	 SSO  2.3~m	  &  DBS      &    4.8  &   $2 \times 900^b$  \\
2003 11 21    &  52964.25     &    73.1     &	 VLT  Antu	  &  FORS1    &     11.5     &   $2 \times 300^c$\\
2003 11 28    &  52971.69     &    80.5     &	 SSO  2.3~m	  &  DBS      &     4.8  &   $2 \times 1200^b$\\ 
2003 12 28    &  53001.53     &   110.3     &	 SSO  2.3~m	  &  DBS         &  4.8  &   $2 \times 1200^b$ \\
2004 01 30    &  53034.45     &   143.2     &	 SSO  2.3~m	  &  DBS         &  4.8  &   $2 \times 1200^b$\\ 
2004 07 25    &  53211.37     &   320.2     &	 VLT  Kueyen	  &  FORS1       & 11.5  &   $4 \times 1200^c$\\ 
\hline
\end{tabular} \\
\begin{tabular}{lll}
$^a$In all tables, ``phase'' refers to days past $B_{\rm max}$, which
  occurred at MJD = $52891.20 \pm 0.30$.\\
$^b$One exposure in the blue arm and one in the red arm. \\
$^c$Exposures shared between the 300V and the 300I grisms.\\
\end{tabular}
\end{table*}

\subsection{Cerro Tololo observations}

Optical photometry of SN~2003hv was obtained with the 
Cassegrain Direct Imager attached to the 0.9-m 
telescope located at the Cerro Tololo Inter-American Observatory (CTIO).
Ten epochs of $UBVRI$ imaging were obtained from 1 to 74 days past 
$B_{\rm max}$. 
 In addition, the CTIO 1.3-m telescope equipped with ANDICAM  was used to collect 15 epochs of early-phase $YJHK_s$ photometry that covers the flux evolution 
from 1 to 62 days past $B_{\rm max}$.

\subsection{KAIT observations}

The 0.76-m Katzman Automatic Imaging Telescope \citep[KAIT;][]{2001ASPC..246..121F}
observed SN~2003hv in $BVRI$ over the course of 42
epochs. These observations were obtained beginning 1 day after $B_{\rm
  max}$ and extend to $+$135 days.

\subsection{Las Campanas observations}

Thirteen epochs of optical photometry covering 10 to 109 days past $B$
maximum were obtained with the Swope 1-m telescope at Las Campanas
Observatory (LCO) during the course of the Carnegie Type II Supernova
(CATS) program \citep{Hamuy2009}.  These images were obtained with a
set of $UBVRI$ filters and a SITe3 detector.  Four optical spectra
were obtained at LCO from 1 day to 17 days past maximum.  A journal of
the spectroscopic observations is provided in Table
\ref{tab:spectralog}.

\subsection{Siding Spring observations}

The European Supernova Collaboration \citep[ESC; see,
e.g.,][]{2004MNRAS.348..261B}, as part of a European Research Training
Network on the study of SNe~Ia,\footnote{\url{www.mpa-garching.mpg.de/~rtn}}  
obtained four optical
spectra and five epochs of photometry at the Siding Spring Observatory
(SSO). These data were collected with the Imager and the
Double Beam Spectrograph (DBS) on the 2.3-m telescope. 
They cover the intermediate phases from 52 to 202 days past maximum.

\subsection{VLT observations}

The Very Large Telescope (VLT) was primarily used to obtain late-epoch
observations in the optical and NIR.  With the exception of one epoch
of optical imaging and spectroscopy obtained on day $+73$, the VLT
observations were conducted from day $+$320 to $+$767. This portion of
the study complements and extends our previous studies of SN 2000cx
(S04) and SN 2001el (SS07) by over 200 days. $U$- and $K_{s}$-band
observations, however, were only carried out to day $\sim$540.
  
Optical imaging was performed with the FOcal Reducer and low
dispersion Spectrographs (FORS1 and FORS2) mounted on the Antu (VLT --
UT1) and Kueyen (VLT -- UT2) telescopes.  These observations were
obtained in service mode under favorable conditions. Details can be
found in Table~\ref{tab:VLTopticallog}.  The optical data can be 
divided into three main epochs: approximately 340, 520, and 700
days past maximum light. No $B$-band imaging was conducted during the 
middle epoch.

Near-infrared imaging ($J_{s}HK_{s}$) was performed with ISAAC
(Infrared Spectrometer And Array Camera) mounted on the Antu
telescope.  Imaging was obtained in the short-wavelength mode using a
jitter-offset technique.  To facilitate a multi-wavelength study, the
NIR images were taken at epochs similar to those of the optical images.
A log is presented in Table~\ref{tab:VLTIRLOG}.

On day $+$320 a nebular spectrum was taken with FORS1. 
Exposures were obtained with the 300$V$ and 300$I$ grisms, and
an order-separation filter (OG590) was used with the latter grism.
The nebular spectrum of SN~2003hv was obtained using a $1\farcs3$
slit and $2 \times 20$ min exposures per grism. The final 
wavelength range is 3600--9750~\AA. 

\subsection{HST observations}

Late-phase observations in the optical were acquired with {\it HST}
during two main epochs, at 307 and 433 days past maximum
brightness. These data were collected as part of a Snapshot Survey of
the sites of Nearby Supernovae.\footnote{Program GO-10272, PI
  Filippenko.} The observations were taken with the Advanced Camera
for Surveys (ACS) using the HRC detector and the F435W, F555W, F625W,
and F814W filters.

A dedicated {\it HST} programme was approved to probe the very late
phases of SN~2003hv.\footnote{Program GO-10513, PI Milne.}
Unfortunately, the optical-band observations, scheduled on day $+$816,
were never executed due to a technical problem with acquiring a guide
star.  Deep NIR observations with filter F160W (similar to the
$H$-band filter) were obtained with the Near Infrared Camera and
Multi-Object Spectrometer (NICMOS) on day $+$786.  The {\it HST}
observations are summarized in Table~\ref{tab:HSTlog}.

%%%%%%%%%%%%%%%%%%%%%%%%%%%%% DATA REDUCTIONS   %%%%%%%%%%%%%%%%%%%%%%%%%%%%%%%%%%%%%%%%

\section{Data reductions}
\label{sec:red}

\subsection{Optical photometry}

\begin{figure}
\includegraphics[width=\columnwidth,clip=]{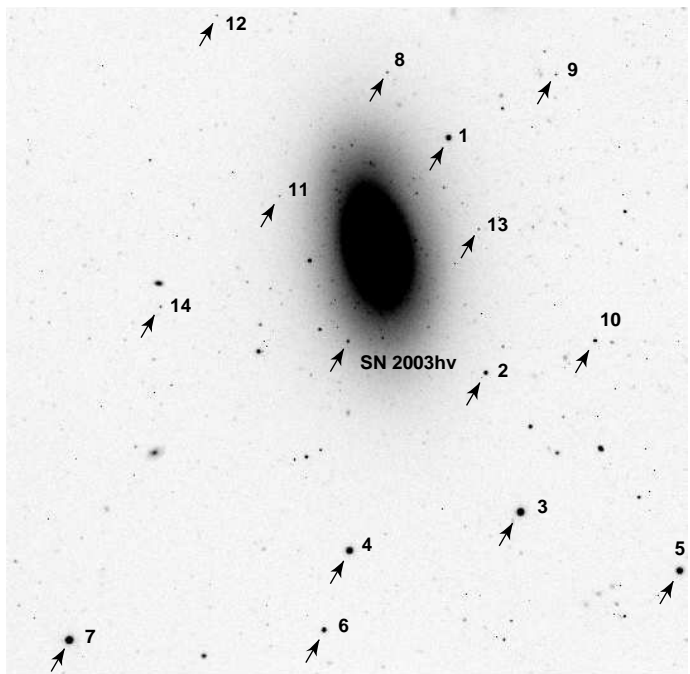}
\caption{The galaxy NGC~1201 including SN~2003hv. The field of view is
  6.8$\arcmin$ $\times$ 6.8$\arcmin$; north is up and east to the
  left. This $B$-band image was obtained on 12 August 2004 with
  VLT/FORS1, when the SN was almost 340 days past maximum. The local
  comparison stars are indicated with arrows. Their calibrated
  magnitudes are listed in Table~\ref{tab:opticalsequence}. Stars 1--7
  are brighter and were used for differential photometry in images
  obtained with smaller telescopes, when the SN was bright. Stars
  8--14 were used mainly at later times.}
\label{locseq}
\end{figure}

All images were reduced in a standard manner including bias and
flatfield corrections using IRAF scripts.\footnote{IRAF is distributed
  by the National Optical Astronomy Observatory (NOAO):
  \url{http://iraf.noao.edu/iraf/web/}} The ground-based photometry of
SN~2003hv was measured differentially with respect to a calibrated
sequence of local stars in the field of NGC~1201.  Given the wide
range in the SN brightness over the different epochs ($>$12 mag), it
was necessary to calibrate both relatively bright and faint local
sequence stars.  The former were used to measure differential
photometry of the SN on the CTIO, KAIT, LCO, and SSO images, while the
latter were used with the late-epoch VLT observations.  The comparison
stars were calibrated with the use of standard-star observations;
  they are indicated in Fig. \ref{locseq}, and their magnitudes are
  listed in Table~\ref{tab:opticalsequence}.   The brightest
  stars used for the early epochs (i.e., stars 1--7) were calibrated
  with the help of five sets of observations on three different
  photometric nights.
To calibrate the faint local sequence, the standard-star field
PG0231+051 \citep{1992AJ....104..340L} was observed under photometric
conditions on 12 August 2004 with the VLT.  Instrumental magnitudes
of the local stars were measured with a $1\farcs6$ aperture radius (8
pixels) and then an aperture correction was applied.  The associated
uncertainties were computed by adding in quadrature the errors
associated with the nightly zero-point, the photometry error as
computed by the IRAF task {\tt phot}, and the error associated with
the aperture correction.

\begin{table*}[tbh]
\caption{Calibrated magnitudes for the local comparison stars in 
Fig.~\ref{locseq}.$^a$}
\label{tab:opticalsequence}
\centering
\begin{tabular}{cccccc}
\hline\hline
Star & $U$ & $B$ & $V$ & $R$ & $I$ \\
\hline
1     & 18.228(0.002) & 18.000(0 .013) & 17.261(0.014) & 16.849(0.013) & 16.431(0.022)\\
2     & 19.712(0.329) & 19.254(0.024) & 18.071(0.019) & 17.323(0.008) & 16.644(0.019)\\
3     & 17.308(0.009) & 16.195(0.009) & 14.919(0.010) & 14.123(0.010) & 13.431(0.008)\\
4     & 17.879(0.006) & 16.738(0.013) & 15.502(0.009) & 14.731(0.010) & 14.062(0.008)\\
5     &  \nodata      & 16.978(0.012) & 16.428(0.016) & 16.093(0.016) & 15.766(0.005) \\
6     &  \nodata      & 18.612(0.014) & 17.779(0.008) & 17.253(0.008) & 16.797(0.012)\\
7     &  \nodata      & 16.228(0.003) & 15.754(0.002) & 15.432(0.002) & 15.123(0.003) \\
8     & 22.179(0.059) & 22.029(0.015) & 21.290(0.017) & 20.847(0.025) & 20.470(0.045)  \\
9     & 21.801(0.058) & 21.665(0.011) & 20.925(0.016) & 20.459(0.023) & 20.043(0.042)  \\
10    & 19.522(0.057) & 20.123(0.010) & 19.933(0.015) & 19.586(0.023) & 19.064(0.042)  \\
11    & 23.595(0.080) & 22.479(0.026) & 21.008(0.019) & 20.069(0.024) & 19.069(0.043)  \\
12    & 23.606(0.071) & 22.399(0.014) & 20.894(0.016) & 19.930(0.023) & 19.018(0.042)  \\
13    & 22.393(0.060) & 22.206(0.018) & 21.352(0.019) & 20.822(0.025) & 20.410(0.045)  \\
14    & 22.384(0.060) & 21.850(0.013) & 20.885(0.016) & 20.302(0.023) & 19.834(0.042)  \\
\hline
\end{tabular} \\
\begin{tabular}{lll}
$^a$Numbers in parentheses are uncertainties.\\
\end{tabular}
\end{table*}

The host galaxy at the location of the supernova was subtracted from
all early-epoch science images (i.e., CTIO, LCO, SSO, KAIT) with the
aid of host-galaxy templates. The template images were obtained with
KAIT and the LCO duPont (+Tek5) 2.5-m telescope at times which were
sufficiently late that the supernova was no longer detectable.  Next,
point-spread function (PSF) photometry of the local sequence and the
supernova was computed from the template-subtracted images in the
manner described by \citet{2006PASP..118....2H}.  Galaxy subtraction
was not performed on the late-time images because the templates were
not deep enough.  At late times (i.e., VLT), when the SN was faint,
aperture photometry with small apertures ($\approx0\farcs8$) was used
to extract the light from the supernova  and the comparison
  stars. This was done to minimize background contamination from the
host galaxy.  PSF photometry was also performed (with tasks in the
{\tt daophot} package) and good agreement was found with the 
aperture-photometry results.

Photometry of the {\it HST} images was done following the procedures
described in the ACS Data Handbook \citep{acshandbook} and in
\citet{2005PASP..117.1049S}. The flux of the supernova was measured
directly in the drizzled images with a 7-pixel aperture ($0\farcs175$)
and converted to the Vega magnitude system using the zero-points
provided by the STScI
webpage.\footnote{\url{www.stsci.edu/hst/acs/analysis/zeropoints}}
Aperture corrections as described by \citet{2005PASP..117.1049S} and a
small charge transfer efficiency (CTE) correction were also applied.

\subsection{S-Corrections}

It is well known that combining data obtained with different
telescopes can lead to systematic differences in the light curves of
SNe.  This was the case for SN~2003hv as well, with differences that
were most pronounced in the $R$ and $I$ light curves around the time
of the secondary $I$-band maximum.  To remedy this problem we computed
and applied S-corrections to our photometry
\citep{2002AJ....124.2100S}.  This practice has become a standard
procedure in recent years and several authors have obtained
encouraging results \citep[see,
  e.g.,][]{2007A&A...469..645S,2007MNRAS.376.1301P,2008MNRAS.388..971P,2009ApJ...697..380W}.
 In the case of SN~2003hv the number of optical spectra available,
  especially at phases up to $+$50 days, does not offer the desired
  temporal and wavelength coverage necessary to compute accurate
  S-corrections.  For this reason a different approach was chosen
  and our S-corrections were computed up to day $+70$ based on the
  spectral template sequence of \cite{2007ApJ...663.1187H}.  At each
  epoch, the spectral template was multiplied by a smooth spline such
  that synthetic photometry constructed with this ``warped'' spectrum matched
  the observed photometry of SN~2003hv.  This is a common practice
  when using SN spectral templates to compute K-corrections
  \citep{2002PASP..114..803N,2007ApJ...663.1187H}.  

Armed with the modified template spectra, standard procedures to
  compute S-corrections were followed; as a reference we adopted the
  $BVRI$ filter transmission functions from
  \citet{1990PASP..102.1181B}, modified to account for the 
  photon-counting nature of CCD detectors. 
The response functions used for KAIT and the CTIO 0.9-m telescope have
been published by \citet{2009ApJ...697..380W} and the ones for SSO
2.3-m telescope by \cite{2007MNRAS.376.1301P}.  We constructed the
corresponding response functions for the LCO Swope telescope by multiplying 
the filter transmission functions with the detector quantum efficiency,
the mirror reflectivity, and the atmospheric transmission, as functions
of wavelength.  To test the accuracy of the modeled passbands, they
were used to derive synthetic magnitudes of a set of
spectrophotometric standards \citep{2005PASP..117..810S}.  These
magnitudes were then used to derive color terms, which were compared
to the color terms derived from the broad-band photometry.  From this
exercise we confirmed that our modeled passbands, in general,
reflected a reliable model of the true global response function of
each telescope.  Finally, in order to compute the S-corrections we
convolved the instrumental bandpasses with the spectral template
sequence by \cite{2007ApJ...663.1187H} modified, as described
  above, to match the SN~2003hv photometry.  No corrections were
attempted either for the $U$ band, or for epochs past day $+$70,
since this is where the spectral template sequence of
\cite{2007ApJ...663.1187H} ends.

\begin{table}[tbh]
\caption{S-corrections (mag) for phases up to $+$70 days.$^a$ }
\label{tab:S-corrections}
\centering
\begin{tabular}{crrrr}
\hline\hline
MJD & $B$  & $V$ & $R$ & $I$ \\
\hline
52892.40  &    0.010	&   0.022   &  0.014	&   $-$0.034	  \\
52892.47  &    $-$0.003	&    0.002    &   0.036   &    $-$0.017	  \\
52893.39  &    0.010	&   0.021   &  0.015	&   $-$0.033	  \\
52893.41  &    $-$0.004	&    0.003    &   0.036   &    $-$0.018	  \\
52894.37  &    0.010	&   0.020   &  0.016	&   $-$0.032	  \\
52895.37  &    0.011	&   0.017   &  0.017	&   $-$0.031	  \\
52896.38  &    0.013	&   0.013   &  0.018	&   $-$0.035	  \\
52896.52  &    $-$0.005	&    0.007    &   0.037   &    $-$0.018	  \\
52897.37  &    0.015	&   0.009   &  0.019	&   $-$0.038	  \\
52897.51  &    $-$0.005	&    0.008    &   0.036   &    $-$0.018	  \\
52898.52  &    $-$0.006	&    0.009    &   0.034   &    $-$0.018	  \\
52899.52  &    $-$0.006	&    0.010    &   0.032   &    $-$0.018	  \\
52901.40  &    0.000	&   0.013   &  0.014	&   0.029	  \\
52901.44  &    $-$0.010	&    0.010    &   0.027   &    $-$0.019	  \\
52902.30  &    $-$0.004	&    0.014    &   0.014   &    0.036	  \\
52902.48  &    $-$0.016	&    0.010    &   0.025   &    $-$0.021	  \\
52903.46  &    $-$0.020	&    0.009    &   0.024   &    $-$0.022	  \\
52904.49  &    $-$0.022	&    0.009    &   0.021   &    $-$0.023	  \\
52905.26  &    0.034	&   $-$0.009    &   0.007   &    $-$0.054	  \\
52905.30  &    $-$0.012	&    0.015    &   0.010   &    0.057	  \\
52906.40  &    $-$0.011	&    0.016    &   0.009   &    0.066	  \\
52906.48  &    $-$0.018	&    0.008    &   0.018   &    $-$0.025	  \\
52907.30  &    $-$0.011	&    0.016    &   0.008   &    0.072	  \\
52908.30  &    $-$0.012	&    0.016    &   0.008   &    0.078	  \\
52910.47  &    $-$0.020	&    0.006    &   0.012   &    $-$0.026	  \\
52912.47  &    $-$0.023	&    0.007    &   0.013   &    $-$0.024	  \\
52914.40  &    $-$0.015	&    0.015    &   0.004   &    0.084	  \\
52914.46  &    $-$0.028	&    0.009    &   0.015   &    $-$0.020	  \\
52916.46  &    $-$0.024	&    0.011    &   0.017   &    $-$0.018	  \\
52919.42  &    $-$0.014	&    0.015    &   0.018   &    $-$0.014	  \\
52921.40  &    $-$0.012	&    0.017    &   0.019   &    $-$0.010	  \\
52925.43  &    $-$0.011	&    0.019    &   0.020   &    $-$0.007	  \\
52928.42  &    $-$0.015	&    0.019    &   0.024   &    $-$0.007	  \\
52929.40  &    $-$0.004	&    0.022    &   0.005   &    0.096	  \\
52930.42  &    $-$0.020	&    0.018    &   0.028   &    $-$0.007	  \\
52932.37  &    $-$0.021	&    0.018    &   0.030   &    $-$0.006	  \\
52934.41  &    $-$0.019	&    0.018    &   0.031   &    $-$0.005	  \\
52936.40  &    $-$0.015	&    0.018    &   0.030   &    $-$0.005	  \\
52939.36  &    $-$0.011	&    0.018    &   0.029   &    $-$0.004	  \\
52942.36  &    $-$0.007	&    0.017    &   0.026   &    $-$0.002	  \\
52942.66  &    $-$0.000	&    $-$0.007   &    $-$0.004  &	0.046	  \\
52945.38  &    $-$0.005	&    0.017    &   0.024   &    $-$0.000	  \\
52948.32  &    $-$0.004	&    0.016    &   0.023   &    $-$0.002	  \\
52951.30  &    0.004	&   0.016   &  0.004	&   0.084	  \\
52954.36  &    $-$0.003	&    0.016    &   0.022   &    $-$0.004	  \\
52959.33  &    0.004	&   0.016   &  0.019	&   $-$0.004	  \\
52961.28  &    0.022	&   0.003   &  0.004	&   0.010	  \\
  \hline
  \end{tabular} 
\begin{tabular}{lll}
$^a$We have used these corrections everywhere in this paper by adding\\
  them to the corresponding values of Table~\ref{tab:phot}.
\end{tabular}
\end{table}

The resulting values are listed in Table~\ref{tab:S-corrections}.
 Due to the fact that these S-corrections were not computed in the
  optimal way (i.e., based on a well-sampled series of real SN~2003hv
  spectra), we have chosen to provide the {\em uncorrected} photometry
  of SN~2003hv in Table~\ref{tab:phot}.  It is left to any future user
  of these data to decide whether to make use of these S-corrections.  
  An estimate of their accuracy can be based on the dispersion
  of the difference between the corrections derived from the templates
  of \cite{2007ApJ...663.1187H} and real spectra. In most cases, this
  error was estimated to be small ($<$0.01 mag), but in a few cases the
  uncertainty in the correction was of the same order as the
  correction itself. This illustrates the limitations of the adopted 
  method.  Our experience, however, showed that the S-corrections did
  reduce the scatter in the light curves, and we have therefore chosen
  to use them everywhere throughout this paper and recommend their
  use.  Nevertheless, some inconsistencies between the various data
sets still remain, especially at times of $+$100--140 days.
Resolving these remaining discrepancies is beyond the scope
of this paper.

\subsection{Near-IR photometry}

The NIR standards P9106 and P9172 \citep{1998AJ....116.2475P},
  were observed with the CTIO 1.3-m telescope on six photometric
  nights, in order to calibrate field stars in the vicinity of
  SN~2003hv.  $Y$-band magnitudes of the Persson stars were calculated
  using the following relationship \citep{2004AJ....127.1664K},
  derived from synthetic photometry of Vega, Sirius, and the Sun: $(Y
  - K_s) = -0.013 + 1.614 (J_s - K_s)$.
Stars 2 and 3 (Fig.~\ref{locseq}) were calibrated in this manner;
their magnitudes are listed in Table~\ref{tab:NIRsequence}.  These
stars also have $JHK$ values in the Two Micron All Sky Survey (2MASS)
and good agreement was found between our values and 2MASS.  The NIR
photometry of the SN was then calculated differentially with respect
to star 3 (the brightest) for all CTIO 1.3-m epochs.
Table~\ref{tab:VLTnirPhot} contains the derived photometry.  

Reductions of the VLT NIR images were done with the {\tt Eclipse}
software package \citep{1999ASPC..172..333D}.  The task {\tt jitter}
was used to estimate and remove the sky background from each
individual image before creating a stacked image.  Photometry of the
supernova was determined on the reduced images relative to  the
  two field stars mentioned above and a third 2MASS star in the field
  of view of ISAAC.  Instrumental magnitudes were computed with {\tt
  phot} using an aperture with a radius of $0\farcs5$.  The quoted
uncertainties in Table~\ref{tab:VLTnirPhot} account for the {\tt phot}
error, the scatter around the zero-point, and the minimum error of the
2MASS sequence.  In the cases when two observations were obtained
during the same night, they were combined to increase the
signal-to-noise ratio.  For the final epoch, images from different
nights were also combined. However, since there was no detection of
the supernova during the final epoch, a 3$\sigma$ upper limit was
computed.

For the {\it HST} NICMOS data, we used the {\it Mosaic Files}
\citep[see][]{nicmosDataHandbook} generated by the STScI pipeline.
The 16 dithered frames were combined in IRAF.  Unfortunately, the SN
was located only 5 pixels away from the erratic NICMOS middle
column\footnote{\url{www.stsci.edu/hst/nicmos/performance/anomalies}},
which proved difficult to correct.  For this reason, and the fact that
the SN was faint, photometry of the SN was performed with a small 
2.5-pixel ($0\farcs1875$) aperture.  For photometry performed on half of
the frames (the ones where due to the dithering pattern the SN
was located farther from the bad column), a larger aperture was used
and these magnitudes were found to be consistent with the results
obtained with the small aperture.  The encircled flux was computed
using the zero-points provided on the STScI
webpage.\footnote{\url{www.stsci.edu/hst/nicmos/performance/photometry}}
Appropriate aperture corrections were applied
\citep[][Fig. 4.10]{nicmosInstrumenthandbook}, and the flux was
converted to the Vega-based magnitude system.

\subsection{Spectroscopy}

All spectra were reduced following standard procedures.
The two-dimensional 
frames were bias subtracted and divided by a master flatfield image, 
and then the cosmic rays were removed.
One-dimensional spectra were extracted, 
wavelength calibrated with comparison-lamp spectra,
and then flux-calibrated relative to spectrophotometric standard star 
observations.
The wavelength calibration was checked against the night-sky lines, and when 
appropriate the blue and red spectra were combined to create a final spectrum. 
Spectra obtained at SSO had telluric features removed with the aid 
of a telluric standard star.

\setcounter{table}{4}

\begin{table*}[tbh]
\caption{Calibrated NIR magnitudes for local comparison stars in Fig.~\ref{locseq}.}
\label{tab:NIRsequence}
\centering
\begin{tabular}{ccccc}
\hline\hline
Star & $Y$$^a$ & $J$ & $H$ & $K_s$  \\
\hline
2     & 16.103 (0.017)  & 15.792 (0.011) &  15.295 (0.011) &   14.927 (0.031) \\
3     & 12.842 (0.013)  & 12.501 (0.009)  & 11.868 (0.013)  &  11.732 (0.017) \\
\hline
\end{tabular} \\
\begin{tabular}{lll}
$^a$There can be an associated systematic uncertainty of 0.03 mag in
  the $Y$-band calibration \citep{2004AJ....127.1664K}. \\
\end{tabular}
\end{table*}

\begin{table*}[tbh]
\caption{Near-infrared photometry of SN~2003hv.}
\label{tab:VLTnirPhot}
\centering
\begin{tabular}{ccclcccc}
\hline\hline
Date & MJD$^a$ &Phase & Telescope & $Y$ & $J$$^b$ & $H$ & $K_{s}$ \\
(UT)  &  (days) & (days)    &                      &(mag)   &(mag)  &  (mag)   &  (mag)      \\
\hline
2003 09 10  &   52892.37  &  1.2   &  CTIO 1.3~m  &  13.252(0.015)   &  13.081(0.015)   &  13.286(0.015)  &   13.100(0.018)   \\
2003 09 14  &   52896.28  &  5.1   &  CTIO 1.3~m   &  13.591(0.015)   &  13.792(0.015)   &  13.423(0.015)  &   13.378(0.021)   \\
2003 09 17  &   52899.28  &  8.1   &  CTIO 1.3~m   &  13.716(0.015)   &  14.435(0.015)   &  13.502(0.015)  &   13.529(0.026)   \\
2003 09 24  &   52906.31  & 15.1   &  CTIO 1.3~m   &  13.403(0.015)   &  14.491(0.015)   &  13.202(0.015)  &   13.211(0.018)   \\
2003 09 28  &   52910.26  & 19.1   &  CTIO 1.3~m   &  13.031(0.015)   &  14.179(0.015)   &  13.112(0.015)  &   13.189(0.022)   \\
2003 10 02  &   52914.20  & 23.0   &  CTIO 1.3~m   &  12.845(0.016)   &  14.000(0.018)   &  13.283(0.020)  &   13.216(0.073)   \\
2003 10 05  &	52917.25  & 26.1   &  CTIO 1.3~m   &  13.025(0.015)   &  14.306(0.015)   &  13.546(0.015)  &   13.524(0.020)   \\
2003 10 08  &	52920.24  & 29.0   &  CTIO 1.3~m   &  13.255(0.015)   &  14.657(0.015)   &  13.799(0.015)  &   13.796(0.026)   \\
2003 10 11  &	52923.22  & 32.0   &  CTIO 1.3~m   &  13.439(0.015)   &  14.980(0.015)   &  13.958(0.015)  &   13.936(0.022)   \\
2003 10 14  &	52926.25  & 35.1   &  CTIO 1.3~m   &  13.650(0.015)   &  15.236(0.016)   &  14.145(0.016)  &	\nodata      \\
2003 10 17  &	52929.22  & 38.0   &  CTIO 1.3~m   &  13.844(0.015)   &  15.519(0.015)   &  14.273(0.015)  &   14.116(0.026)   \\
2003 10 20  &	52932.17  & 41.0   &  CTIO 1.3~m   &  14.037(0.015)   &  15.725(0.017)   &  14.431(0.016)  &   14.306(0.024)   \\
2003 10 24  &	52936.21  & 45.0   &  CTIO 1.3~m   &  14.468(0.015)   &  16.310(0.023)   &  14.770(0.018)  &	\nodata     \\
2003 11 03  &	52946.21  & 55.0   &  CTIO 1.3~m   &  14.912(0.015)   &  16.790(0.036)   &  15.041(0.020)  &   15.060(0.048)   \\
2003 11 10  &	52953.22  & 62.0   &  CTIO 1.3~m   &  15.207(0.017)   &  17.088(0.065)   &  15.420(0.037)  &	\nodata     \\
2004 08 18  	  & 53235.37 &  344.2 & VLT Antu  & \nodata   &  20.394(0.140)  & \nodata &	 \nodata    \\
2004 08 29   	  & 53246.31 &  355.1 & VLT Antu  & \nodata   &  \nodata  & 19.977(0.106)  &   20.221(0.216)      \\
2004 08 30        & 53247.28 &  356.1 & VLT Antu  & \nodata   & 20.078(0.122)   & \nodata &	20.303(0.224)	\\
2005 01 24        & 53394.06 &  502.9 & VLT Antu  & \nodata   &  \nodata  & 20.163(0.138) &	 $>$ 20.700	  \\
2005 02 24        & 53425.99 &  534.8 & VLT Antu  & \nodata   &  \nodata  & \nodata &    $>$ 20.700	 \\
2005 02 27        & 53428.99 &  537.8 & VLT Antu  & \nodata   &  \nodata  & 20.495(0.146) &	 \nodata    \\
2005 02 28        & 53429.99 &  538.8 & VLT Antu  & \nodata   &  21.087(0.144)  & \nodata &	 \nodata    \\
2005 10 14$^c$    & 53657.07 &  765.9 & VLT Antu  & \nodata   &  $>$ 22.310  & $>$ 22.000 &	 \nodata    \\
2005 11 03  	  & 53677.31 &  786.1 & {\it HST} NICMOS$^d$ & \nodata   &  \nodata  & 22.691(0.051) &	 \nodata    \\ 
\hline
\end{tabular} \\
\begin{tabular}{lll}
$^a$All images obtained the same night are referred to a mean MJD \\
$^b$A $J$ filter was used at the CTIO 1.3~m telescope, while $J_s$ was used at the VLT.\\
$^c$All images obtained on October 13--15  are averaged here. \\
$^d$Vega magnitudes in the {\it HST} filter system ($F160W$).\\
\end{tabular}
\end{table*}

%%%%%%%%%%%%%%%%%%%%%%%%%%%%% RESULTS   %%%%%%%%%%%%%%%%%%%%%%%%%%%%%%%%%%%%%%%%
\begin{figure*}[th]
\includegraphics[width=\textwidth,clip=]{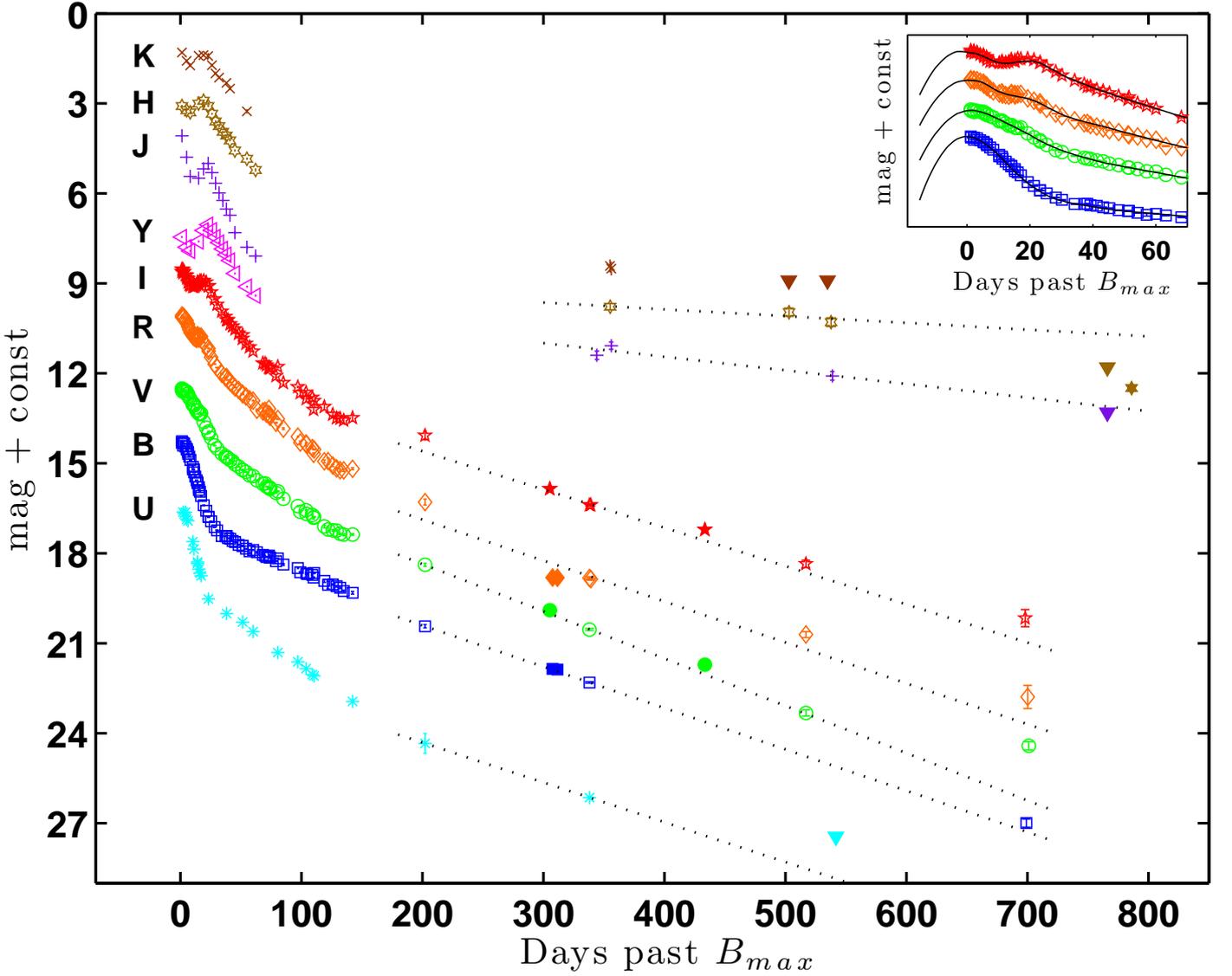}
\caption{The $UBVRIYJHK$ light curves of SN~2003hv.  They have been
  shifted for clarity by the following constants: $+$4.5, $+$1.8, 0.0,
  $-$2.4, $-$4.2, $-$5.8, $-$9, $-$10.2, and $-$11.8, respectively.
  Filled triangles represent 3$\sigma$ upper limits.  The {\it HST} points
  (in the {\it HST} filter system) are indicated with filled symbols.
  Linear fits to the data in the range $+$200--540 days are shown with
  dotted lines to guide the eye.  The inset
  contains light-curve fits to the early $BVRI$ photometry
  \citep{Burns2009}; the data have been shifted vertically by different
  constants than in the main graph.}
\label{AllLC}
\end{figure*}

\section{Results}
\label{sec:res}

\begin{figure}[tbh]
\includegraphics[width=\columnwidth,clip=]{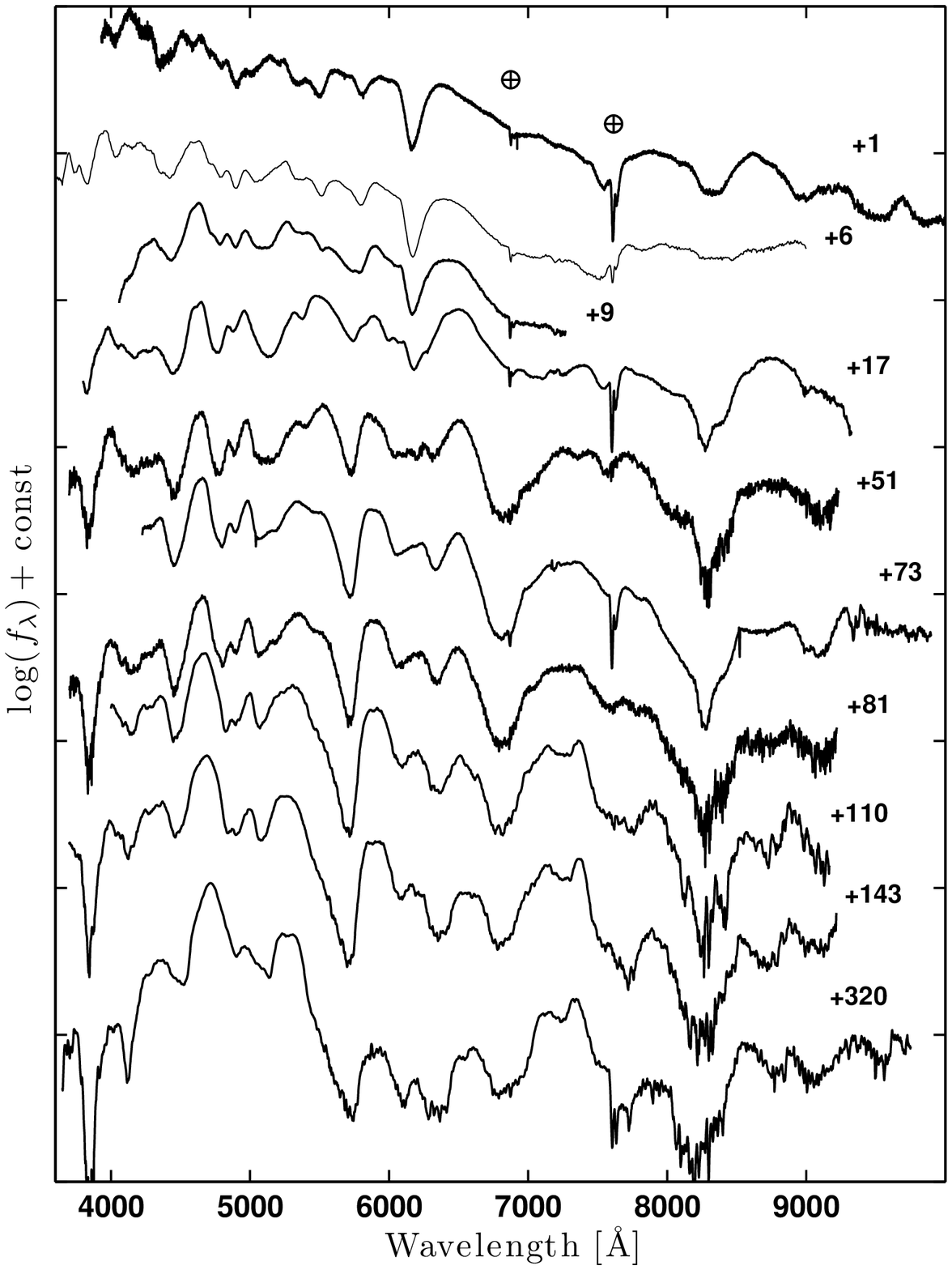}
\caption{
Spectral evolution of SN~2003hv from maximum light to the nebular phase.  
Numbers indicate days past maximum brightness. For clarity, the spectra
have been offset in flux scale with respect to each other.  The Earth
symbols mark telluric features. Note that the SSO spectra have had
their telluric lines removed.  The bottom three spectra have been
smoothed (by a moving average of 5 pixels) for presentation purposes.}
\label{spectroscopy}
\end{figure}

We now present the results of our observations, first for the early
phases and then for the later epochs. The complete light curves and
spectral sequence for SN 2003hv are shown in Figs.~\ref{AllLC} and
\ref{spectroscopy}, respectively.

\subsection{Early-phase photometry}
\label{sec:earlyphot}

Using the method described by \citet{2006ApJ...647..501P}, we fit
template light curves to the early-time S-corrected $BVRI$ photometry.
This allowed us to determine basic parameters of the light curves and
to estimate the level of host-galaxy reddening.  Our slightly modified
fitting method is described by \citet{Burns2009}.  The fits can be
seen in the inset of Fig.~\ref{AllLC}.

The light-curve fit indicates that $B_{\rm max}$ occurred on MJD $=$
52891.2$\pm$0.3 (or 9.15 September 2003) with an apparent peak
magnitude of 12.45$\pm$0.03.  Peak brightness in the $VRI$ bands
occurred at $+$0.9, $+$0.1, and $-$2.2 days relative to $B_{\rm max}$.
 We note here that the reported uncertainties are the statistical
  errors of the fit. Since our photometry does not cover the peak,
  however, a larger systematic uncertainty might be anticipated.

The computed host-galaxy reddening is $E(B-V)_{\rm host} = -0.04 \pm
0.01$ mag.  Formally, one must also add a systematic uncertainty of
$\pm0.06$ mag, related to the observed color spread of SNe~Ia.  We note
that zero reddening is consistent with the position of the supernova
in the outskirts of an S0 galaxy.

The light-curve fit of SN~2003hv yields a $B$-band decline rate
$\Delta m_{15}(B) = 1.61 \pm 0.02$ mag.  SN~2003hv therefore lies near
the faint end of the normal luminosity vs. decline-rate distribution,
close to the highly subluminous SN~1991bg-like SNe~Ia
\citep[e.g.,][]{1992AJ....104.1543F}, which are observed to have
$\Delta m_{15}(B)> 1.7$ mag.  To date there is a lack of well-studied
SNe~Ia with $\Delta m_{15}(B)$ values in the range between 1.5 and 1.7
\citep[see, e.g.,][]{2006ApJ...647..501P, 2007Sci...315..825M}.
Currently it is not clear to what extent the properties of normal and
subluminous SNe~Ia compare.  In this sense, SN~2003hv, with its
extensive data coverage, may help us understand the similarities or
differences between these different types of SNe~Ia.

The distance modulus of SN~2003hv is estimated to be $\mu_{\rm SN} =$
31.58 $\pm$ 0.05 mag from these light-curve fits.  The quoted error is
the statistical error of the fit and we have assumed a Hubble constant
of $H_0=$ 72 km s$^{-1}$ Mpc$^{-1}$ \citep{2001ApJ...553...47F}.
There is an additional systematic error (amounting up to $\sim$0.15
mag) related to the intrinsic dispersion in the luminosity of SNe~Ia.
This distance is consistent with the SBF distance of NGC 1201
$\mu_{\rm SBF} = 31.37 \pm 0.30$ mag
\citep{2001ApJ...546..681T,2003ApJ...583..712J}.  We conclude that
SN~2003hv was a normal-luminosity SN Ia that obeys the Phillips
relation \citep{1993ApJ...413L.105P,1999AJ....118.1766P} within its
inherent scatter, since this is an underlying assumption for deriving
a distance modulus with the aid of the light curves.  Throughout the
rest of the paper, in order to avoid any circular reasoning, we adopt
the $\mu_{\rm SBF}$ distance to NGC~1201.  The conclusion that
SN~2003hv is a normal SN~Ia is also
supported in part by the existence of a secondary maximum in the $I$
band, whereas very rapidly declining SN~1991bg-like events lack this
feature \citep{1992AJ....104.1543F}.

By fitting only the KAIT data, which is a sufficiently large
homogeneous dataset, no significant differences were obtained for the
derived light-curve parameters. However, the scatter around the fit
was reduced ($\chi_{\rm dof}^2$ of 3.9 versus 9.3), since this avoids
systematic uncertainties introduced by combining data from different
telescopes. The $\chi_{\rm dof}^2$ for the combined dataset without
S-corrections was 15.6.

\begin{figure}
\includegraphics[width=\columnwidth,clip=]{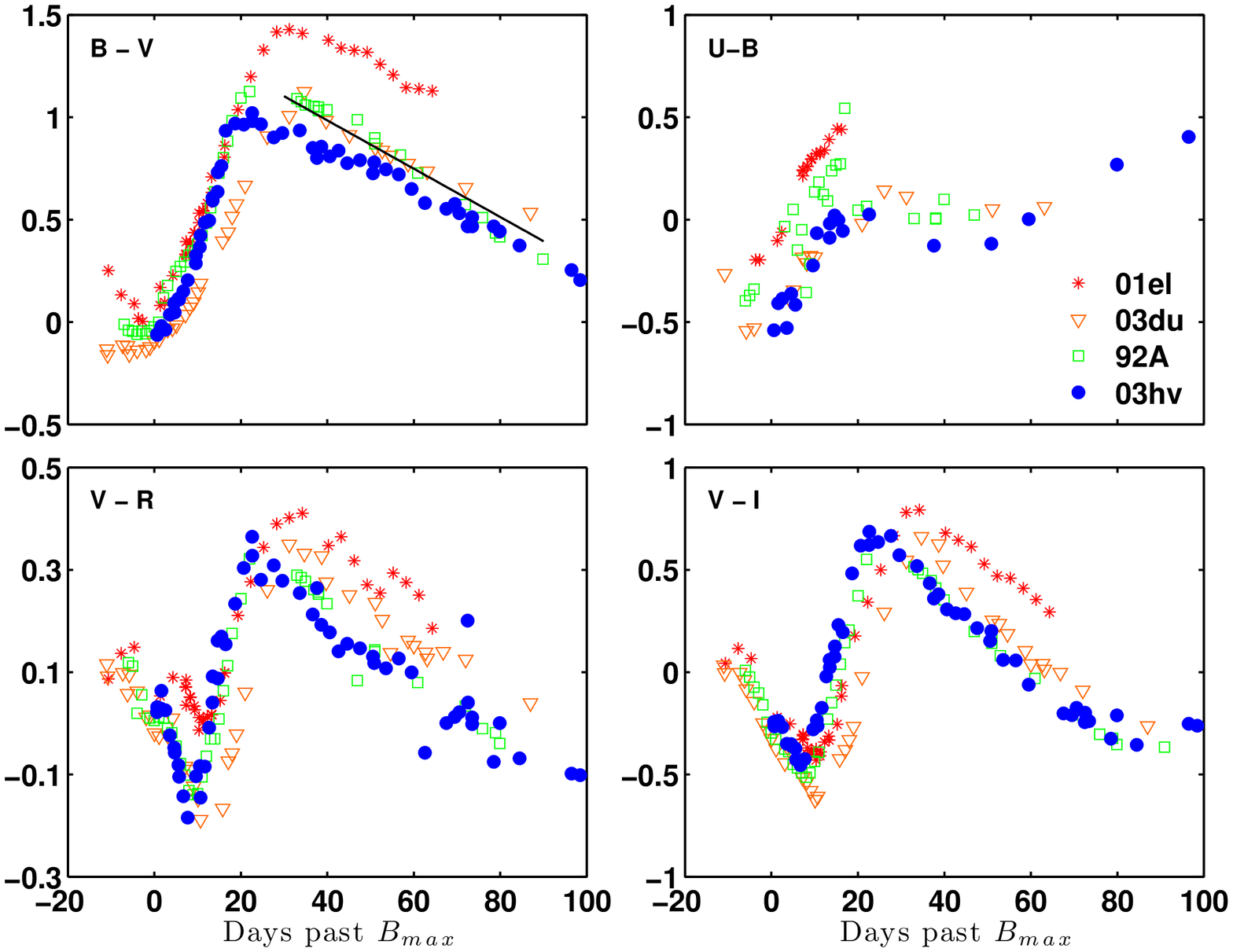}
\caption{ 
Color evolution of SN 2003hv from maximum light to 100 days past
maximum. For comparison, we have plotted the colors of SN~1992A
\citep{1996ssr..conf...41S}, SN~2001el \citep{2003AJ....125..166K},
and SN~2003du \citep{2007A&A...469..645S}. The Phillips/Lira relation
\citep{1999AJ....118.1766P} is also indicated in the $B-V$ panel
(solid line). The four SNe have only been corrected for Galactic
reddening \citep[$E(B-V) = 0.016$, 0.017, 0.014, and 0.010 mag,
  respectively,][]{1998ApJ...500..525S}. This illustrates that
SN~2003hv, as in the cases of SNe~1992A and 2003du, suffers little to
no extinction from its host. SN~2001el, on the other hand, was
substantially reddened and clearly does not follow the Phillips/Lira
relation.}
\label{color}
\end{figure}

In Fig.~\ref{color}, the early observed colors of SN~2003hv are
compared with those of the normal SN~1992A
\citep{1996ssr..conf...41S}, SN~2001el \citep{2003AJ....125..166K},
and SN~2003du \citep{2007A&A...469..645S}.  These SNe~Ia have $\Delta
m_{15}(B)$ values of 1.47, 1.15, and 1.06, respectively.  The $V-R$
and $V-I$ colors of SN~2003hv are similar to those of SN~1992A (the
one with the most similar decline rate), while the $B-V$ color starts
slightly bluer around maximum (similar to SN~2003du) and then three
weeks later matches that of SN~1992A.  The $B-V$ color evolution
between 30 and 90 days past maximum follows the Phillips/Lira relation
\citep{1999AJ....118.1766P}.  Note that in Fig.~\ref{color}, the
colors were only corrected for Galactic extinction to illustrate that
SN~2003hv is similar to SNe having little or no host-galaxy reddening
(SNe~1992A and 2003du).  This is additional evidence that SN~2003hv
was a normal SN~Ia that suffered negligible host-galaxy extinction.

\subsection{Early-phase spectroscopy}
\label{sec:earlyspec}
 
Figure~\ref{spectroscopy} displays the spectroscopic sequence from 1
to 320 days past maximum.  The sequence consists of four early-phase,
five ``mid-epoch,'' and one late-phase spectra.

The earliest spectra display intermediate-mass elements characteristic
of a normal SN~Ia near maximum brightness. The evolution of these
spectra confirms that this was a normal event.  We have used the
``Supernova Identification'' code \citep[SNID;][]{2007ApJ...666.1024B}
to compare the early-time spectra of SN~2003hv with a library of
supernova templates.  SNID indicates a good agreement ({\it r}lap
quality parameter values $>$ 10) with several normal SNe~Ia (e.g.,
SNe~1992A,~1994D, and 1996X) at epochs similar to those deduced from
our light-curve fits, to within a few days.

At maximum light, the ratio of the depth of the \ion{Si}{ii}
$\lambda$5972 and $\lambda$6355 absorption features was 0.40 $\pm$
0.05.  This value is somewhat larger than what is measured in other
normal SNe~Ia (SN~1992A being the runner-up with 0.38), but smaller
than values found for subluminous SN 1991bg-like events
\citep{1995ApJ...455L.147N,2004ApJ...613.1120G,2005ApJ...623.1011B}.

From the four early-epoch spectra, the rate of decrease in the
expansion velocity of the \ion{Si}{ii} $\lambda$6355 feature was
measured to be $\dot{v} =$ 41$\pm$6 km~s$^{-1}$~day$^{-1}$. By
comparing with a large sample of events, we find that SN~2003hv lies
in the low-velocity gradient (LVG) group as defined by
\citet{2005ApJ...623.1011B}.

\subsection{Late-phase Photometry}
\label{subsec:latephot}

The measured decay rates of the late-time light curves are listed in
Table~\ref{tab:decratespaper}.  To facilitate comparison with results
of previous studies \citep[][S04; SS07]{2006AJ....132.2024L}, these
decline rates were computed as linear slopes at time intervals out to
$\sim$540 days.  The measured late-phase $BVR$-band decline rate of
$\sim$1.4 mag per 100 days in SN~2003hv is similar to values reported
for other well-studied SNe~Ia.  A possible exception is the $I$ band,
which is found to decline slower than $BVR$, but not as slowly as
found in other SNe~Ia (i.e., $\sim$1.0~mag per 100 days).  In
addition, as discussed by SS07, the $U$ band has a late-time decline
rate comparable to that of $BVR$ for at least up to 340 days.

However, in the case of SN~2003hv, the evolution of the late-epoch
light curve is, for the first time, followed in multiple bands out to
700 days past maximum light.  At these very late stages, an apparent
slowdown of the optical light curves is observed; the drop in
luminosity appears to be decelerating.  This behavior is most
pronounced in the $V$ band, where it has also been observed in several
other SNe~Ia.  We discuss this further in
Sect.~\ref{subsec:slowdown}.

\begin{figure*}
\sidecaption
\includegraphics[width=12cm]{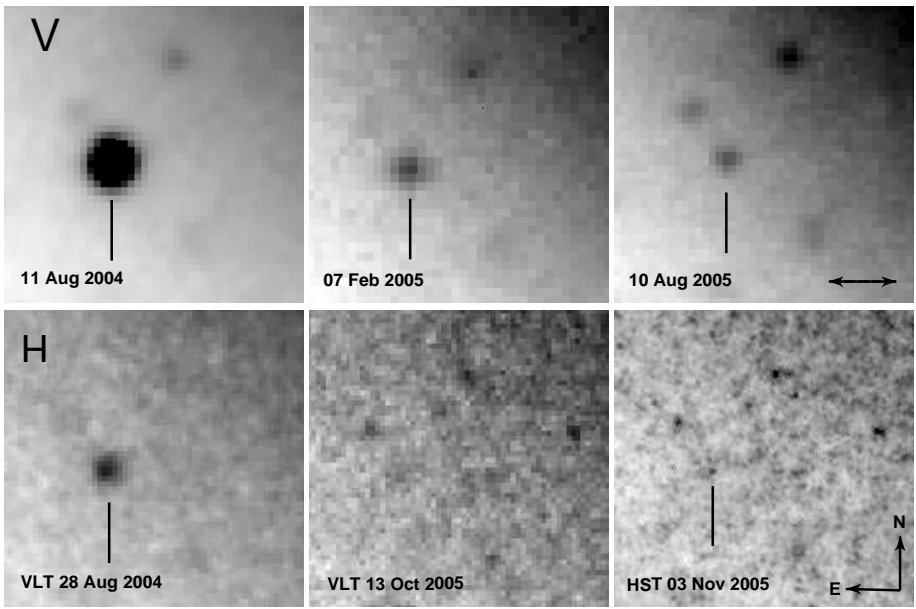}
\caption{
The fading of SN~2003hv between days $+$340 and $+$786 as observed in
the $V$ band (upper sequence) and the $H$ band (lower sequence).  The
double arrow in the last $V$-band frame is 2$\arcsec$ across.
SN~2003hv is clearly detected in all $V$-band images.  It is not
significantly detected in the VLT/ISAAC $H$-band image from 13 October
2005, at an upper limit of 22 mag.  However, SN~2003hv is detected in
a deeper image obtained 3 weeks later with {\it HST}/NICMOS on 786 days past
maximum with a magnitude of 22.69 $\pm$ 0.05.  This is the latest-ever
detection of a SN~Ia in the NIR.  The last image is constructed from
the drizzled combination of 16 frames.  }
\label{VHfade}
\end{figure*}

In the NIR bands, SN~2003hv displays a different behavior than the
constant brightness observed during late-phase observations of SNe
2000cx and 2001el (S04 and SS07, respectively).  SN~2003hv slowly
fades in brightness between days $+$350 and $+$540.  Furthermore, the
SN is not detected in the VLT images at $+$766 days, indicating a
further drop during this period.  However, we do detect a point-like
source in the {\it HST}/NICMOS/F160W image obtained at $+$786 days
(see Fig.~\ref{VHfade}).  From astrometry relative to nearby stars
with the help of the deep VLT $H$-band images, we establish that this
object's position is consistent with the position of the supernova to
within 0$\farcs$061 $\pm$ 0$\farcs$085 (where the quoted error is the
transformation error).  With a magnitude of 22.69 $\pm$ 0.05, this is
to our knowledge the latest detection ever achieved of a SN~Ia in the
NIR.

\begin{table*}
\centering
\caption{Decline rates in mag per 100 days.$^a$ }
\label{tab:decratespaper}
\begin{tabular}{lccccccccc}
\hline\hline
Epoch  &  $U$ & $B$ & $V$ & $R$ &$I$ &$J_s$ &$H$ &$K_s$ \\
\hline\hline
200--540 days$^b$  &    1.33(0.24) &  1.38(0.05)  &  1.58(0.03)  &  1.37(0.05)  &   1.28(0.06) & 0.45(0.09)	& 0.23(0.09) &$>$0.30  \\	
\hline
\end{tabular}\\
\begin{tabular}{l}
$^a$Numbers in brackets are the formal errors of the weighted 
least-squares linear fits to the photometric data.\\
$^b$In the $U$ band only to 340 days.\\
\end{tabular}
\end{table*}

Notice that the {\it HST} ACS observations were not included in the
linear fits in Table~\ref{tab:decratespaper}.  This is because the
{\it HST} and ground-based filter systems are significantly different.
These differences can produce systematic effects in the photometry,
especially when one considers the nonstellar nature of the spectral
energy distribution of SNe~Ia.  Synthetic photometry computed with the
nebular spectrum and the VLT and {\it HST} filter transmission
functions indicates that on day $+$320 the SN would appear brighter by
0.20, 0.25, and 0.28 mag in $BVI$ and fainter by 0.37 mag in $R$ in
the {\it HST} filter system.\footnote{The F625W is actually an $r$,
  not an $R$, filter.}  However, the farther we move from the epoch of
the nebular spectrum, the less accurate these corrections become;
hence, they were not applied.  Nevertheless, the ACS points lie fairly
close to the calculated linear slopes.

The $R$-band $+$516 day observation was obtained using the FORS2
``special'' $R$-band filter, which differs from the Bessell $R$
filter.  Synthetic photometry computed with the nebular spectrum
reveals that SN~2003hv would appear $\sim$0.08 mag brighter in the
$R_{\rm special}$ filter at $+$320 days.  Assuming that the shape of
the spectrum does not change out to $+$516 days, this induces a change
to the respective $R$-band decline rates by no more than half a
standard deviation, and it has no consequence to the following
discussion.

\subsection{Late-phase spectroscopy}
\label{subsec:latespec}

The nebular spectra (bottom Fig.~\ref{spectroscopy}) of SN~2003hv are,
generally speaking, similar to those of other normal SNe~Ia, and
display several broad iron-group emission features.  In particular,
the latest spectrum is dominated by strong Fe emission features at
$\sim4700$~\AA\ and 5250~\AA. These are produced from the blending of
forbidden transitions associated with \ion{Fe}{ii} and \ion{Fe}{iii}.

\begin{figure}
\includegraphics[width=\columnwidth,clip=]{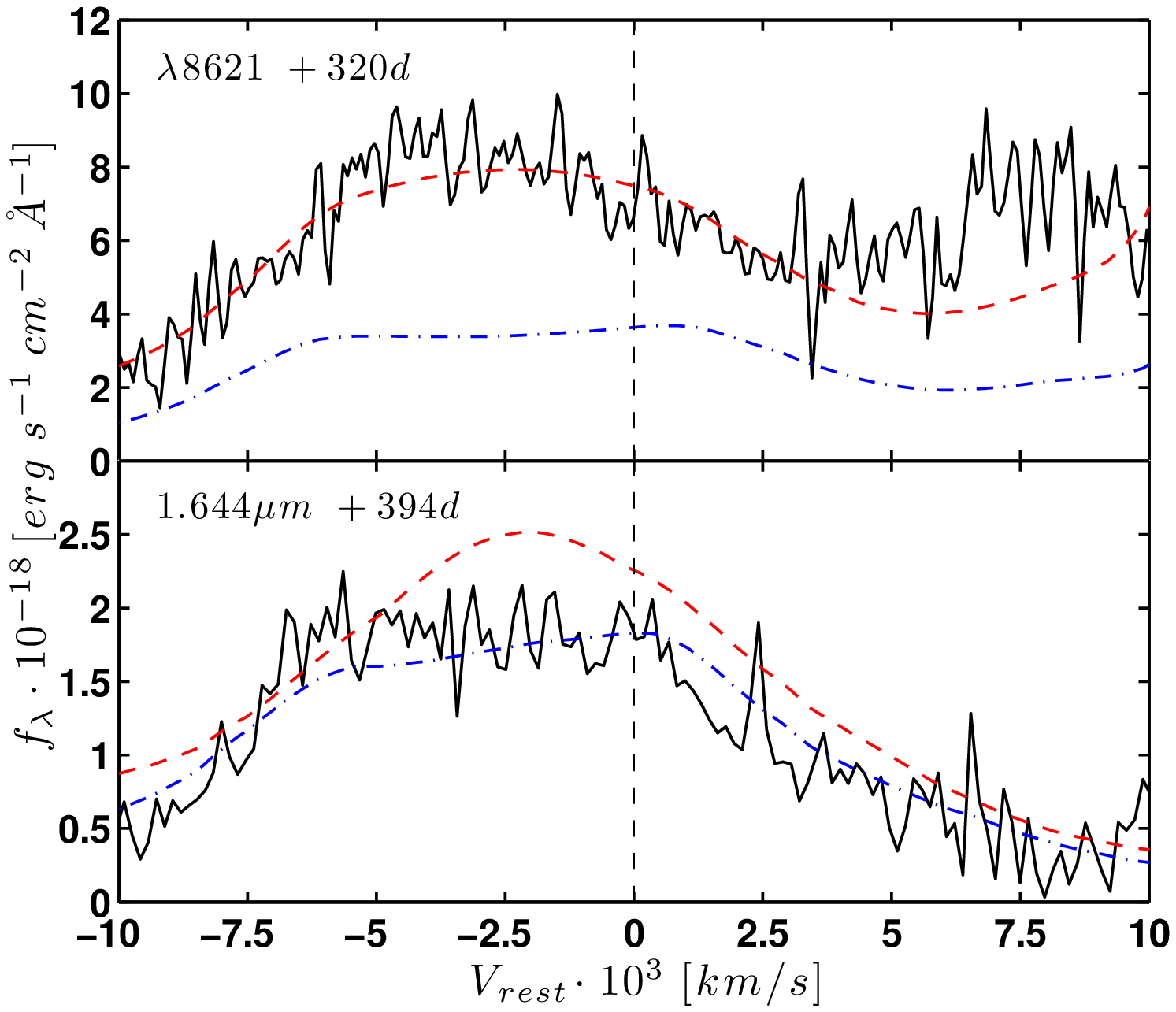}
\caption{
Comparison of the [\ion{Fe}{ii}] line profiles at 8621~\AA\ (our
spectrum) and 1.644 $\mu$m \citep{2006ApJ...652L.101M} in velocity
space.  These are relatively clean, isolated [\ion{Fe}{ii}] lines
according to our model.  The two lines appear equally blueshifted, and
their blueshift is equal to the blueshift of [Co~III] 11.89 $\mu$m
\citep[see][]{2006ApJ...652L.101M,2007ApJ...661..995G}.  The 1.644
$\mu$m line exhibits a flat-topped profile while the 8621~\AA\ line is
not clearly ``peaked'' (assuming it is not a result of blending by
neighboring lines).  Also shown are the model spectra (without
photoionization) at $+$300 days (red dashed line) and at $+$400 days
(blue dashed-dotted line) blueshifted by 2600 km s$^{-1}$.  In the
top panel, their flux has been scaled down by a factor of 7.5 so that
the $+$300-day model matches approximately the observed flux. The same
has been done in the bottom panel, matching the $+$400-day model by
scaling it down by a factor of 10.  Notice that the 8621~\AA\ line is
also expected to develop a clear flat-topped profile at later phases.
}
\label{lineVelComp}
\end{figure}

\citet{2006ApJ...652L.101M} and \citet{2007ApJ...661..995G} found in
the NIR and MIR nebular spectra of SN~2003hv that the [\ion{Fe}{ii}]
1.644~$\mu$m and [\ion{Co}{iii}] 11.89~$\mu$m emission lines were
blueshifted by $\sim$2600~km~s$^{-1}$.  These shifts were interpreted
as the result of an asymmetric, off-centre explosion.  In addition,
the [\ion{Fe}{ii}] 1.644~$\mu$m line had a flat-topped profile.  It
was therefore argued \citep{2006ApJ...652L.101M} that regions below
$\sim$3000~km~s$^{-1}$ are filled with neutron-rich non-radioactive
isotopes \citep[see also][]{2004ApJ...617.1258H}.

We examined our optical nebular spectrum to investigate whether any
signatures of blueshifted lines or flat-topped profiles were
present. It should be stressed that the optical spectrum is produced
from the blend of many overlapping transitions, and it is therefore
problematic to make strong claims from the appearance of optical
lines.  However, a fairly isolated [\ion{Fe}{ii}] feature is present
at 8621~\AA.  This line, which often falls outside the wavelength
coverage of optical spectra, contributes according to our modeling
(see Sect.~\ref{subsec:fullspec}) $\sim$90\% of the flux in this
wavelength region.  This is comparable to the ``cleanness'' of the
[\ion{Fe}{ii}] 1.644~$\mu$m line in the NIR and higher than the 65\%
contribution that our model suggests for the [\ion{Fe}{ii}]
$\lambda$7155 line in its corresponding region. The latter line, which
is contaminated mainly by [\ion{Ca}{ii}], was inspected by
\cite{2004ApJ...617.1258H} in the case of SN~2003du, and was found to
have a peak ``seemingly in contradiction'' with the boxy NIR profiles.
In Fig.~\ref{lineVelComp} the 8621~\AA\ and 1.644~$\mu$m
\citep{2006ApJ...652L.101M} lines are compared in velocity space.  We
see that these two features have somewhat similar profiles and both
are apparently blueshifted by $\sim$2600 km~s$^{-1}$. The same could
be argued for the 7155~\AA\ line.  Other iron lines are more
heavily blended and definitely unsuitable for such diagnostics (notice,
however, that the 5250~\AA\ feature is also flat-topped).  This
could be considered as supporting the findings presented by
\citet{2006ApJ...652L.101M} and \citet{2007ApJ...661..995G}. However,
in addition to the concerns raised above, these lines do show some
evolution (Fig.~\ref{spectroscopy}), which could be additional
evidence pointing toward blending.

%%%%%%%%%%%%%%%%%%%%%%%%%%%%%   DISCUSSION    %%%%%%%%%%%%%%%%%%%%%%%%%%%%%%%%%%%%%%%%%

\section{Discussion}
\label{sec:disc}

\subsection{SN 2003hv in the context of its $\Delta m_{15}(B)$ value}

In this subsection, the properties of SN~2003hv are reviewed in the
light of its rather uncommon light-curve decline-rate parameter.
There is a general division between normal and subluminous SNe~Ia,
where the latter are typically observed to have $\Delta m_{15}(B) >
1.7$~mag.  The subluminous SNe~Ia do not seem to follow the linear
luminosity vs. decline-rate relation of normal SNe~Ia
\citep{1993ApJ...413L.105P,2006ApJ...647..501P}, although an
exponential fit might be able to include them
\citep[e.g.,][]{2004ApJ...613.1120G}.  It is not clear whether
subluminous SNe~Ia can be related to a different class of progenitors
or explosion mechanisms \citep[see, e.g.,][and references
  therein]{2000ARA&A..38..191H}.

As very few SNe~Ia with 1.5 $< \Delta m_{15}(B) <$ 1.7 have been
observed (and none in as much detail as SN~2003hv), we feel it is
warranted to discuss SN~2003hv in this context, even if it is just a
single example, and confirm that it is a ``normal'' member of the SN~Ia
family.  In particular, with respect to several relations proposed for
SNe~Ia, we make the following remarks.

\labelitemiii~ 
Adopting the independent SBF distance measurement, an absolute
magnitude of $B_{\rm max} = -18.99 \pm 0.35$ is deduced for
SN~2003hv, which is fully consistent with the expected luminosity from
its decline rate \citep[linear fits, by][give
  $-$19.0]{2006ApJ...647..501P}.

\labelitemiii~  
Adding the SN~2003hv $\cal R (\ion{Si}{ii})$ value (0.40 $\pm$ 0.05)
to the $\cal R (\ion{Si}{ii})$ vs. $\Delta m_{15}(B)$ correlation
\citep{2004ApJ...613.1120G,2005ApJ...623.1011B}, we find that this
data point nicely connects the normal SNe~Ia to the subluminous group
in a previously unexplored area of this relation.

\labelitemiii~  
\citet{1998ApJ...499L..49M} showed that the full width at half-maximum
intensity (FWHM) of the line at 4700~\AA\ in nebular spectra of SNe~Ia
correlates well with $\Delta m_{15}(B)$.  Although SN~2003hv has a
FWHM value slightly larger than the one expected by the correlation in
\citet{1998ApJ...499L..49M}, it is consistent with the existence of
such a correlation.

\labelitemiii~  
\citet{2001ApJ...559.1019M} suggested the cut between the normal and
subluminous SNe~Ia to occur at $\Delta m_{15} = 1.6$ mag as far as the
late light-curve shape is concerned.  However, this was based on very
few SNe in the appropriate range and the study was later updated with
the inclusion of SN~1999by \citep{2005coex.conf..183M}.  We note here
that in this sense the late-time light curve evolution of SN~2003hv is
intermediate between SN~1992A and the transitional (with respect to
late-time light curve behavior) SN~1986G.

 \labelitemiii~ 
The peak luminosity of SN~2003hv in the NIR bands is comparable to those of normal SNe~Ia. 
As pointed out by \citet{Krisciunas2009}, this is also the case for  a sub-sample of SNe~Ia 
with $\Delta m_{15}(B) > 1.6$, that peak in the $J$ band before the $B$ band.
Although our observations do not seem to cover the $JHK$ peaks, 
this is  almost certainly the case for SN~2003hv 
and estimates based on the templates by  \citet{2004AJ....127.1664K} 
give $-$18.52, $-$18.17 and $-$18.33 ($\pm$ 0.31) for $M_J$, $M_H$ and $M_K$, respectively.
The assertion of normal peak brightness does not change even if the observed maxima are used as lower limits, instead of extrapolating back in time.

\labelitemiii~ 
We would not expect to see a SN with such a $\Delta m_{15}(B)$ in a
late-type galaxy
\citep{2000AJ....120.1479H,2007ApJ...659..122J}. Indeed, its presence
in an S0 galaxy is not surprising.

The main conclusion is that SN~2003hv appears to obey many of the
known correlations with respect to the $B$-band decline-rate relation
and is an object that is similar to the other normal SNe~Ia used to
derive these correlations.  Seen from another perspective, we show
that these particular correlations also hold for this previously
underexplored value of $\Delta m_{15}(B)$.

\subsection{Nebular spectrum synthesis}
\label{subsec:fullspec}

A unique aspect of SN~2003hv is our broad wavelength coverage at
nebular phases, when including the infrared spectra from
\citet{2006ApJ...652L.101M} and \citet{2007ApJ...661..995G}.  This
facilitates comparison with model spectra in an unprecedented manner.
We have used our detailed spectral synthesis code \citep[described in
detail by][S04]{1998ApJ...496..946K,1998ApJ...497..431K,2005A&A...437..983K}
to generate a nebular spectrum of SN~2003hv and compare it with the
observational data.  The spectral synthesis code includes a
self-consistent ionization and level population model to calculate the
emission from each radial zone of the ejecta.  Nonthermal excitation
and ionization by gamma rays and positrons are included, as well as
time-dependent effects. Nonlocal scattering of the emission, however,
is not included.  The present calculations use the W7 hydrodynamical
model \citep{1984ApJ...286..644N} as input.

\begin{figure*}
\includegraphics[width=\textwidth,clip=]{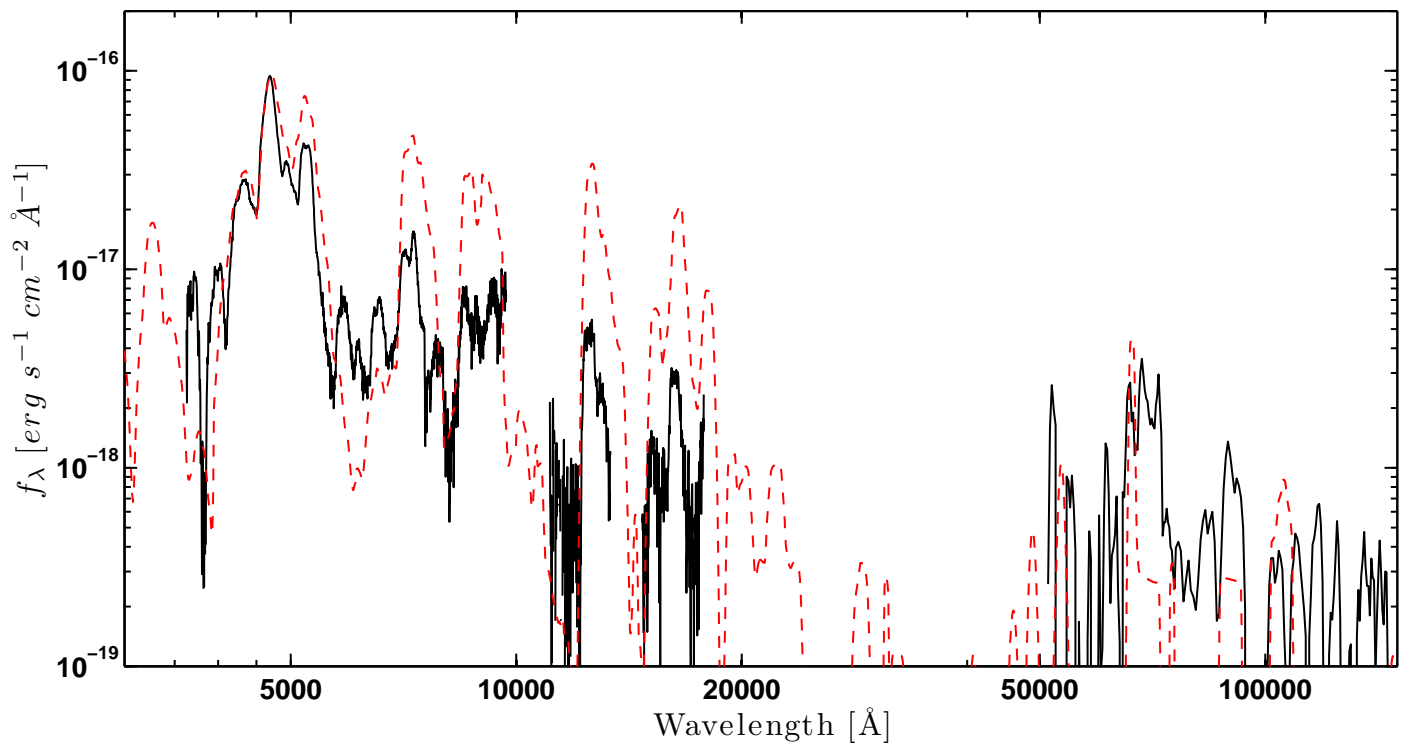}
\caption{
The multi-wavelength nebular spectrum of SN~2003hv (solid line).  This
spectrum is a compilation of our optical spectrum at $+$320 days, the
NIR spectrum of \citet{2006ApJ...652L.101M} at $+$394 days, and the
{\em Spitzer} MIR spectrum of \citet{2007ApJ...661..995G} at $+$358
days.  The optical and NIR spectra have had their flux scaled to match
the age of the MIR spectrum with the aid of the $V$-band and $H$-band
photometry (see text).  For presentation purposes each spectrum was
smoothed by a moving average of 3 pixels. Note also that the MIR
spectrum has large associated error bars that are not shown here.  The
dashed red line shows our model spectrum (without photoionization) at
$+$400 days.  }
\label{MultiSpectrum}
\end{figure*}

Since the nebular spectra of SN~2003hv were all obtained at slightly
different epochs, the optical and NIR spectra were scaled in flux with
the aid of our photometry to match the date of the MIR spectrum. The
$V$ and $H$ magnitudes were interpolated to day $+$358, assuming the
linear slopes from Table~\ref{tab:decratespaper}.  The final combined
spectrum is shown in Fig.~\ref{MultiSpectrum} together with our model
spectrum at day $+$400.

The synthetic spectrum does a reasonable job in reproducing the
general features of the observed spectrum.  The dominant {Fe} peaks in
the range 4000--5500~\AA\ are reproduced accurately, as is the shape
of the spectrum at 7000--9000~\AA.  In the NIR, the relative fluxes of
the [\ion{Fe}{ii}] lines agree with the observations, but the absolute
fluxes are overpredicted.  In the MIR, our model instead
underpredicts the flux levels. The model also predicts a strong
stable (non-radioactive) [\ion{Ni}{ii}] line at 6.64~$\mu$m, which is
also seen in the observed spectrum.  [\ion{Ar}~{ii}]~and
[\ion{Co}~{ii}] lines are identified, as mentioned by
\citet{2007ApJ...661..995G}.  Note that our modeling includes two
different ways to treat the effect of the scattering of UV photons and
their subsequent consequences for the photoionization of the metals
(representing two limiting cases): \emph{with} and \emph{without}
photoionization (see, e.g., S04).  Here only the model without
photoionization (i.e., assuming that all UV photons are redistributed
to longer wavelengths in the scattering process) has been plotted
because it is a better fit to the observed spectrum.  The model with
photoionization is not able to reproduce the bluest peak around the
4000~\AA\ bumps, since the relevant [\ion{Fe}{i}] emission is
suppressed in this case.

An interesting feature, as mentioned above, is the flat-topped
profiles observed in the NIR lines.  It has been suggested that a
detailed study of the [\ion{Fe}{ii}] 1.644~$\mu$m NIR feature could be
the cleanest probe of the ejecta kinematics \citep[e.g.,][and
  references therein]{2004ApJ...617.1258H}.
\citet{2006ApJ...652L.101M} discussed that the NIR nebular spectra of
two out of four SNe~Ia exhibit a flat-topped profile in this line,
with SN~2003hv being the strongest case and the one that has been
observed farthest from maximum brightness.  Our model was first used
to investigate how ``clean'' this line is.  It was confirmed that even
though the nebular spectrum consists of a large number of overlapping
lines, within our model \ion{Fe}{ii} appears to dominate the
1.7~$\mu$m region. There are a number of different \ion{Fe}{ii}
transitions that contribute to the feature, but since predominantly
recombination radiation is seen at this phase, uncertainties in the
ionization are less important in modeling the feature. As pointed out
above, we also propose that the [\ion{Fe}{ii}] $\lambda$8621 line is
relatively clean and isolated from other contributing lines.

The scenario favoured by \citet{2006ApJ...652L.101M} and
\citet{2007ApJ...661..995G} to explain the flat-topped line profiles
is an inner ``hole'' of unmixed, neutron rich, nonradioactive,
iron-group elements in the core. These are the products of electron
capture, which takes place in the highest density burning regions.
Such a configuration of chemical elements is observed in one
dimensional (1-D) deflagration explosion models (such as W7), or even
1-D delayed detonation (DDT) models, but is incompatible with current
3-D deflagration models, which predict much mixing \citep[e.g.,][]{2005A&A...432..969R,2007ApJ...668.1132R}.
\citet{2006ApJ...652L.101M} modeled the line and found a slight
asymmetry due to overlapping [Fe~II] 1.664 and 1.667~$\mu$m lines, but
argued that the overall flatness nevertheless suggested no efficient
mixing between the highest density burned region and its surroundings.

Our modeling of the optical-IR spectrum at $+$320 days,
contemporaneous with the optical spectrum, did not show any evidence
for flat-topped line profiles. However, allowing the model to run to
even later phases, the flat-topped line profiles start to develop.  By
400 days past maximum, the line becomes flat-topped
(Fig.~\ref{lineVelComp}), as also noted by
\citet{2006ApJ...652L.101M}.  Note that our model assumes complete and
{\it in situ} deposition of the energy carried by the positrons.
The core of the ejecta is the densest region where both gamma
rays and positrons are most efficiently deposited.  The
evolution from peaked to flat-topped can thus be explained by
the fact that even at $+$300 days, there is still a fair
fraction of energy deposited by gamma rays, and we can follow
these as they are deposited in the central region.  At even
later epochs, however, the energy is provided solely by the
positrons. Since these are not able to penetrate to the central
regions, the absence of radioactive material in the center will
give rise to a flat-topped line profile.  In this sense, the
flat-topped line is consistent with complete and local positron
trapping, providing a diagnostic complementary to the late-time
light curves (see Sect.~\ref{sect:positrons}).

Our optical spectrum is useful for constraining some of the
alternative explanations that have been proposed for the flat-topped
line profile.  The alternative idea that the profile is due to a dusty
core \citep{2006ApJ...652L.101M} can be ruled out because no such
evidence is seen at optical wavelengths. Furthermore, the possibility
mentioned by \citet{2007ApJ...661..995G}, that the hole may be filled
with a large fraction of unburned material at low velocities as
suggested by some 3-D deflagration models
\citep[e.g.][]{2005A&A...432..969R}, is also very unlikely: the
absence of the [O~I] emission at 6300~\AA\ is clearly inconsistent
with such a scenario, as discussed by \citet{2005A&A...437..983K}.
\citet{2007ApJ...661..995G} further mention, but do not favor, the
interaction of the companion star as a possible cause of the
hole. This may indeed be interesting to investigate in light of the
new models by \citet{2008A&A...489..943P}, suggesting the presence of
a hole in the wake of the explosion, but the observable signatures of
such a hole filled by the companion star are still rather unclear.

Finally, an alternative explanation, proposed here, for the lack of emission below 2500 km~s$^{-1}$, could be that an IRC has taken place
in the highest density regions.  
This scenario is further discussed in Sect.~\ref{subsec:IC}.

\subsection{Slowdown of optical decay}
\label{subsec:slowdown}

At very late times, the optical decline rates of SN~2003hv appear to
be slowing down, especially in the $V$ band.  We first examine whether
our photometry could be contaminated by a light echo or a background
(or foreground) source.  Light echoes have been observed in the past
for a handful of SNe~Ia
\citep{1994ApJ...434L..19S,2001ApJ...549L.215C,2008ApJ...677.1060W,2008ApJ...689.1186C}.
However, no evidence of extended structure was observed in our PSF
subtractions, neither in the final epoch nor in the high-resolution
{\it HST} images at $+$430 days.  On the other hand, not all possible
echo geometries can be resolved at this distance. In the
single-scattering approximation \citep{2005MNRAS.357.1161P}, it is
estimated that only echos in intervening clouds at distances greater
than $\sim$75~pc from the SN would be resolved by ACS.  But in
addition, our nebular spectrum shows no evidence of blue continuum and
the final colors are not particularly blue.  SN~2003hv did not explode
in a star-forming region, but in the outskirts of an S0 galaxy where
no dust is expected and there are no signs of host extinction.
Finally, if the latest-epoch photometry were dominated by a coincident
background source, the corrected slope at 200--500 days would become
steeper than those previously observed for SNe~Ia at these phases,
indicating that this is probably not the case.

We therefore believe that the final photometry indeed measures the
supernova light.  Actually, such a slowdown has been previously seen
in several other SNe~Ia observed beyond 600 days past maximum in single
(usually $V$-band) observations: SN 1992A \citep{1997A&A...328..203C},
SN 2000E \citep{2006AJ....132.2024L}, and SN 2000cx
(S04, their Sect.~6.5).  Many of the same considerations apply in the
case of SN~2003hv.  However, our multi-color observations suggest that
this slowdown may not be characteristic of only the $V$ band. This
questions the speculations in S04 about [\ion{Fe}{i}] predominantly
emitting in the $V$ band being the explanation for this behavior.

The main consequence of this observation is that the dramatic IRC
predicted by some models did not occur at these phases.

\subsection{Bolometric light curve and \Nif\ mass}
\label{subsec:bolom}

To construct an UV-optical-NIR (UVOIR) light curve of SN~2003hv, we
used the $UBVRIJHK$ photometry  from Tables~\ref{tab:phot} and
  \ref{tab:VLTnirPhot}, including the S-corrections from
  Table~\ref{tab:S-corrections}.  For missing epochs, the photometry
was interpolated by fitting suitable functions to the data.  At early
times, spline interpolation was used, while at the late nebular
phases, linear fits were initially assumed. However, in some bands,
due to deviations from the linear decay, quadratic or cubic
polynomials gave better fits to the data and were adopted.

\begin{figure}
\includegraphics[width=\columnwidth,clip=]{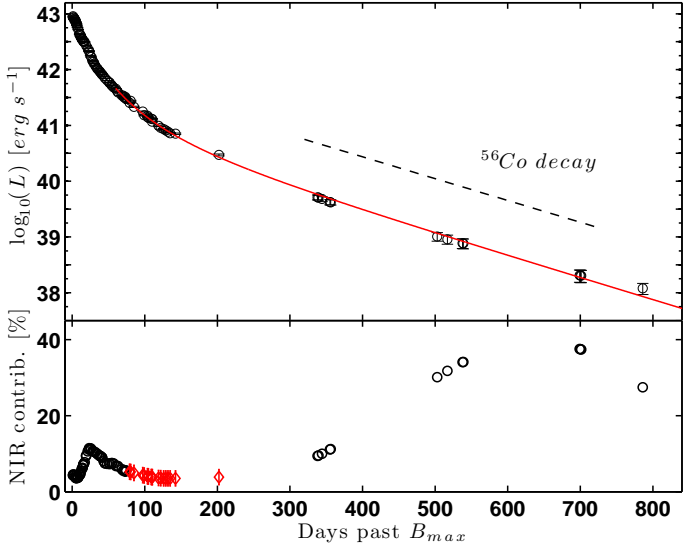}
\caption{  Upper panel: the UVOIR light curve of SN~2003hv
  constructed as described in Sect.~\ref{subsec:bolom}.  The last point
  is based mostly on extrapolations, and should be regarded as rather
  uncertain.  The red solid line is the best fit of a radioactive
  decay energy deposition function of \Cif\ between $+$60 and 700 days.
  The estimated \Nif\ mass from the energy deposition function fit is
   0.22 $\pm$ 0.02 M$_{\sun}$, which is less than what was
  obtained at maximum light (0.40 $\pm$ 0.07 M$_{\sun}$),
  indicating that at these late phases a substantial fraction of the
  flux is emitted outside the UVOIR bands. At 300--700 days past
  maximum the decline rate is linear and follows the expected decay
  time of \Cif\ (dashed line) assuming full and instantaneous
  positron energy deposition.  The displayed error bars do not include
  the error in the distance to NGC~1201.   Lower panel: The
    evolution of the NIR contribution to the UVOIR light curve,
    computed as integrated flux from $J$ to $K$ over the integrated
    UVOIR flux.  Since no NIR data were available at $+$60--200
    days, we assumed a contribution (red diamonds in the graph) that
    maintains the smoothness of the curve and is compatible with the
    corresponding temporal evolution observed by S04. }
\label{bolom}
\end{figure}

In the case of the $U$ band, it was assumed that the light curve
continued the linear decay obtained out to $+$340 days. For the $K$
band we made the limiting assumption that it was \emph{barely} not
detected at $+$534 days, while the further assumption was made that
the $J-H$ and $H-K$ colors do not change between $+$530 and $+$786 days, in
order to estimate $J$ and $K$ magnitudes at the final epoch.  None of
the upper limits in $UJHK$ bands was violated by any of these
assumptions or fits.  The photometry was subsequently corrected for
Galactic extinction assuming $R_V = 3.1$ and following the prescription
of \cite{1998ApJ...500..525S}.  Magnitudes were converted to fluxes
within the individual filters and the UVOIR flux (and its associated
error) was obtained by integrating the filter fluxes over wavelength.
The UVOIR luminosity was calculated assuming a distance of 18.79 $\pm$
2.60 Mpc.  No corrections were applied for the flux lost blueward of
$U$ or redward of $K$. The UVOIR light curve is displayed in
Fig.~\ref{bolom}.

The UVOIR light curve can be used to estimate the amount of
\Nif\ synthesized during the explosion.  Using the estimate of the
peak brightness and application of Arnett's rule
\citep{1982ApJ...253..785A} suggests that 0.40--0.42 M$_{\sun}$
  of \Nif\ was synthesized in the explosion, depending on whether we
  choose to correct by an additional 5\% or 10\% for the flux not
  observed blueward of the $U$ band \citep[][and references
    therein]{2006A&A...450..241S}.  The associated error amounts to
  0.07 M$_{\sun}$, accounting for the errors in the measured flux and
  the uncertainty in the rise time.  The error increases to 0.11
  M$_{\sun}$ if the uncertainty in the distance to NGC~1201 is
  included.

The \Nif\ mass can also be estimated by fitting an energy deposition
function for the radioactive decay of \Cif\ in the tail of the
bolometric light curve.  A simple model, also used by S04 and SS07, is
$L = 1.3 \times 10^{43}M_{\rm Ni}e^{-t/111.3}(1 - 0.966e^{-\tau})$,
where $L$ is the bolometric luminosity, $M_{\rm Ni}$ is the
\Nif\ mass, $t$ is the time past maximum, and $\tau$ is the optical
depth which is taken to evolve as $(t_1/t)^2$, where $t_1$ is the time
where the optical depth to the gamma rays becomes unity
\citep{1998A&A...337..207S}. This model assumes complete and
instantaneous energy deposition from the positrons.  By fitting this
simple equation to our UVOIR light curve in the range $+$60--700 days, we
obtain  $M_{\rm Ni}=$ 0.22 $\pm$ 0.02~M$_{\sun}$ and $t_1=$ 32.7
  $\pm$ 1.8 days.  The quoted errors here are merely the formal
errors from the least-square fits.  Since no $JHK$ observations
  were available at $+$60--200 days, we made an assumption for
  the NIR contribution\footnote{By NIR contribution to UVOIR, we
    define here the ratio of the integrated flux from $J$ through $K$ to
    the integrated flux from $U$ through $K$.} at these phases
  (Fig.~\ref{bolom}, lower panel).  The adopted assumption ensures
  that the evolution of the NIR contribution is smooth and similar to
  that of the well-observed SN~2000cx (S04); that is, we have assumed
  that the NIR contribution continues its smooth decline until it
  reaches a minimum at around $+$130 days, after which there is an
  upturn that slowly leads to the observed high NIR contribution at
  late times.  An associated uncertainty of $\pm$2\% was assumed for
  these calculations.  In addition, it was checked that other
  reasonable assumptions do not change the derived \Nif\ mass by more
  than $\pm$ 0.02 M$_{\sun}$.

We point out that there is a substantial difference between the
\Nif\ mass estimated from the energy emitted in the UVOIR bands
around maximum light compared to that estimated at late phases.  There
could be a number of reasons for this ``missing flux,'' including color
evolution outside the UVOIR bands, positron escape, or an IRC.

Previous studies have shown that color evolution is very important at
these stages within the UVOIR limits: S04 demonstrated that the
contribution of the NIR bands to the UVOIR luminosity increased from
about 3\% to 28\% between $+$150 and $+$500 days. A similar result was
obtained by SS07. We find that this fraction for SN~2003hv evolves
from about 9\% to 30\% between $+$330 and $+$500 days and increases to
$\sim37$\% by $+$700 days (Fig.~\ref{bolom}, lower panel).

S04 and SS07 also estimated that the UVOIR light curve might probe
only $\sim60$\% of the true bolometric luminosity at these late
phases.  Our modeling suggests that at $+$350 days the UVOIR misses
27\% of the total luminosity, while at $+$500 days this ratio
increases to 44\%. It is most likely that the difference in derived
nickel mass is due to a significant fraction of flux at late phases
being emitted in the far-infrared.

By integrating the flux contained within the nebular spectra in
Fig.~\ref{MultiSpectrum}, we find that the flux emitted in the MIR
region probed by {\em Spitzer} is a significant portion of the total
flux probed at day $+$358: $L_{\rm MIR}/L_{\rm tot_{probed}} = (34 \pm 17)$\%.  The large error bar is due to the poor signal-to-noise ratio 
of the MIR spectrum, and there is an additional uncertainty in the 
{\em Spitzer} absolute-flux calibration.  For comparison, our model, for
which the onset of the IRC occurs later (i.e., $\sim$ 500--700 days), 
predicts only a value of 8\% for this ratio. Note also that the
probed MIR range is still blueward of most fine-structure lines in
the far-IR, to which most of the flux would shift in the event
of an IRC.

\subsection{Trapping of positrons}
\label{sect:positrons}

There is an ongoing debate concerning the extent to which the
positrons created in the radioactive decays are trapped in the ejecta.
This discussion has implications for our understanding of the
magnetic field configuration of the ejecta
\citep{1998ApJ...500..360R}. It is of interest for understanding the
Galactic 511 keV emission where SNe~Ia have been suspected to
contribute if the positrons escape
\citep{1999ApJS..124..503M,2008NewAR..52..457P}, and it is also important
in order to properly model the late-time emission of SNe~Ia.  Our
spectral synthesis model assumes that all positrons are trapped and
that they deposit their energy in situ, but more elaborate positron
transfer mechanisms could be envisioned \citep{1999ApJS..124..503M}.

The most straightforward observational test is to measure the
decline rate of the late-time bolometric light curve, since positrons
are the main energy contributors during late phases when the gamma
rays escape the ejecta freely.  In the simplest scenario, complete
positron trapping will result in a late-phase bolometric light curve
that follows the decay rate of radioactive $^{56}$Co (0.98 mag per
100 days), whereas positron escape would produce a faster decay rate.
In particular, one would expect positron escape to become increasingly
important at later phases, and therefore a bolometric light curve that
deviates progressively more from the radioactive input rate
\citep[e.g.,][their Fig.~1]{2001ApJ...559.1019M}.

The late-time UVOIR decline rate of SN~2003hv between $+$300 and
$+$700 days is 0.99 $\pm$ 0.04 mag per 100 days, exactly what is
expected for complete and instantaneous positron trapping
(Fig.~\ref{bolom}). In addition, during these epochs, the UVOIR light
curve does not show large deviations from a linear decay. Only
observations past day $+$300 were included to minimize contamination
from energy deposited by gamma rays, and the data point at $+$786 days
was ignored because it is based mainly on extrapolations (except in
the $H$ band).

This gives little room for energy being lost in the form of positrons,
at least between $+$300 and 700 days. There can, in principle, be
alternative explanations that give a shallower light curve while
allowing for positrons to escape, such as freeze-out
\citep{1993ApJ...408L..25F}.  However, this would require multiple
effects to ``conspire'' to give an extended linear decay with a slope
that perfectly mimics that of radioactive decay.  We therefore believe
that there is no evidence for substantial leakage; it
is adequate to model SNe~Ia at late times assuming that positrons do
not escape.

Turning to the contribution of positrons to the Galactic 511 keV line,
\cite{2008NewAR..52..457P} mentions that a constant escape fraction as
small as 3\% would be enough to make them an important source. It is
difficult to exclude such a small contribution.  We merely note that
as our light curves show little evidence for positron escape at late
phases, it is hard to imagine that positron escape is important at
earlier epochs when the density of the ejecta is considerably higher
and the magnetic field strength is greater.  Of course, SNe~Ia do show
some diversity at late times \citep[e.g., the peculiar
  SN~2006gz;][]{2009ApJ...690.1745M}, but our conclusions seem to hold
for SN~2000cx (S04), SN~2001el (SS07), and SN~2003hv.

We already noted that within the electron capture scenario favored by
\cite{2006ApJ...652L.101M} to explain the flat-topped line profiles,
there is additional evidence for  in situ positron trapping.  If the
positrons from the radioactive isotopes were able to travel inside the
ejecta, they would also excite the central material and not give rise
to a flat-topped profile.

\subsection{Infrared catastrophe}
\label{subsec:IC}

Having discussed the positron trapping and the missing flux at late
times in the UVOIR light curve, we now proceed to the question of
whether an IRC could have occurred in the ejecta of SN~2003hv.

\begin{figure*}
\sidecaption
\includegraphics[width=12cm,clip=]{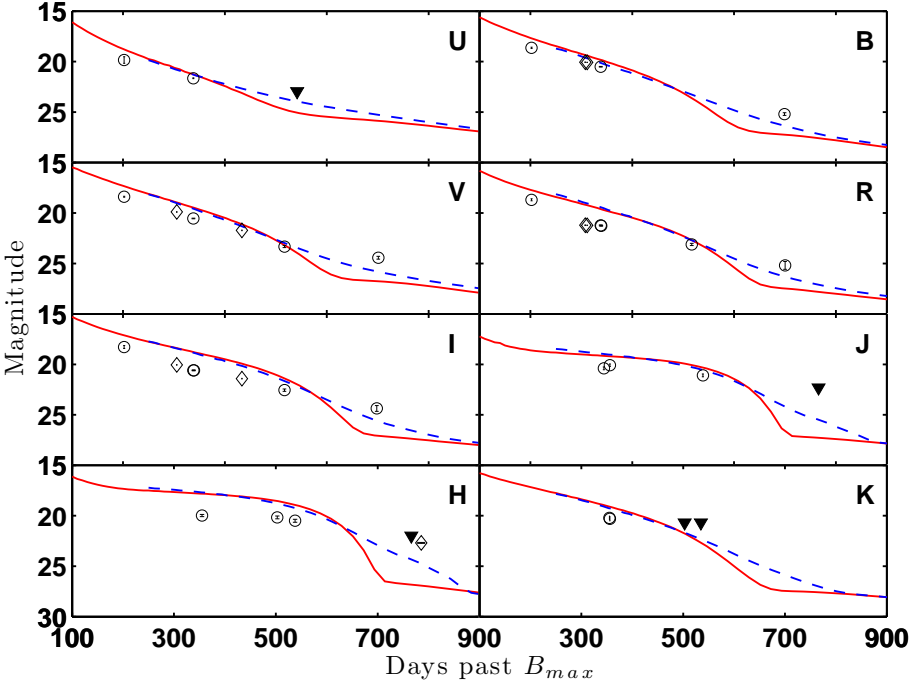}
\caption{Photometric data of SN~2003hv after 200 days past maximum
  compared with the model light curves of W7 and our spectral synthesis
  code (solid red line). Circles are ground-based data, diamonds are
  data from {\it HST}, and filled triangles denote upper limits. The sharp
  drops in the model light curves between $+$550 and $+$700 days are
  due to the predicted IRC.  The observed data do not show such a
  dramatic evolution.  The dashed blue line is a simplified ``clumped''
  model illustrating how clumping can reduce or postpone the effects
  of the IRC: the high-density regions undergo an IRC first, but the
  low-density regions remain hot and continue emitting at the UVOIR
  wavelengths.  }
\label{CecLightCurves}
\end{figure*}

Figure~\ref{CecLightCurves} shows the late $UBVRIJHK$ light curves of
SN~2003hv together with our detailed model light curves based on the W7
model (see also S04). The very sharp drop in the modeled light
curves between $+$550 and 700 days is a consequence of the IRC.

From the modeling point of view, the IRC is expected to occur once the
temperature of the ejecta drops below a critical threshold.  In the
case of SN~2003hv, the observed light curves appear to show little
evidence of such a dramatic scenario.  Not only do the optical bands
seem to demonstrate an opposite trend, as discussed above, but the
$+$786 day $H$-band detection places strong constraints on the drop of
the NIR luminosity. While this has been hinted by single optical
passband observations in the past, this is the first multi-wavelength
study extending to such late phases.

The lack of a sudden and dramatic drop in the late-phase flux suggests
that at least a portion of the ejecta is kept above the critical
temperature limit which marks the onset of the IRC.  One possible
solution that we propose here is clumping of the ejecta. In this case,
lower density regions cool more slowly and remain hot enough to avoid
reaching the IRC temperature limit.  The emission from these regions
may dominate the optical and NIR range, thus causing the flux of the
SN not to decrease as fast as predicted by a model with a more
homogeneous density distribution.  This is illustrated in
Fig.~\ref{CecLightCurves} by the dashed curve, which represents a
model with clumpy ejecta.  In this simplistic model, clumping is
achieved by artificially compressing and decompressing subsequent
layers of the W7 ejecta.  We stress that no effort has been made to
``fit'' the data or to simulate a realistic 3-D situation.  The purpose
of this experiment is to show that clumping can postpone the main
observational signatures of an IRC.  However, even in this model the
high-density regions cool and indeed undergo the IRC.  The lower
density regions, on the other hand, stay hot enough for this not to
occur and therefore continue to emit sufficient flux in the optical
and NIR.

Support for the inhomogeneous nature of the cooling comes from the
calculations by \citet{2005A&A...437..983K}, based on the 3-D model by
\citet{2005A&A...432..969R}: at $+$500 days, a large range of
temperatures (and ionizations) is present, from a few hundred
K to $\gtrsim$6000~K (their Fig.~7). Whether the IRC occurs in a
specific region depends both on the density and composition. The
dependence on composition comes mainly from the difference in
\Nif\ content, which affects the local heating by the positrons.

One way to interpret our observations could thus be to propose that an
IRC {\em has} taken place in the highest density regions of the
ejecta.  With the innermost regions having higher densities, this
scenario can explain the lack of emission in these regions, resulting
in the observed flat-topped line profiles.  This idea can also explain
the missing flux in the tail of the UVOIR light curve as estimated
from the difference of the derived \Nif\ masses at maximum and at late
phases.  It is, however, required that the highest density regions
must have already suffered this IRC before $\sim350$--700 days, since
during these epochs the UVOIR light curve shows little deviation from
the \Cif\ radioactive decay rate of 0.98 mag per 100 days.  The 
less-dense parts still continue emitting in the optical and NIR
wavelengths.  This scenario is also supported by the excess flux
measured in the MIR at $+$358 days.

Further pursuit of the idea of clumpiness is beyond the scope of this
paper, but we suggest this to be an interesting topic of investigation
for future 3-D modeling efforts. Here we restrict ourselves to
pointing out that an IRC might have occurred in the innermost, highest
density regions of SN~2003hv, and that this idea is compatible with
the multi-wavelength observations of the ejecta.

\subsection{Geometry of the explosion}

As also pointed out by \cite{2006ApJ...652L.101M}, the blueshifted
lines observed in some of the emission features of the nebular
spectrum may suggest an asymmetry in the distribution of \Nif\ that
was formed during the explosion.   This is also the conclusion of
  \cite{Maeda2009}, who model the nebular spectra of SN~2003hv with the
  aid of a simplified kinematical model and propose that we are
  viewing the explosion toward an offset high-density region.
\citet{2007A&A...465L..17H} and \citet{2007MNRAS.378....2S} showed
that the viewing angle of off-centre explosions can have a significant
effect on the light-curve properties of SNe~Ia. An enhancement in
brightness is expected in their models when the bulk of \Nif\ is
moving toward the observer. However, this is not observed in
SN~2003hv, which has a normal luminosity.

A different diagnostic of (another kind of) asymmetry comes from the
analysis of spectropolarimetry of SN~2003hv at +6 days (unpublished
data).  While the lack of continuum polarization suggests that the
photosphere was spherical at these times, non-negligible line
polarization (0.19\%) associated with \ion{Fe}{ii} and \ion{Si}{ii}
features implies an asymmetric or clumpy line-forming region for these
two species. Note that the polarization of the \ion{Si}{ii} feature at
these times is at odds with the suggested evolutionary trend of
\ion{Si}{ii} polarization in normal SNe~Ia, for which zero
polarization is expected at $+$6 days after maximum
\citep[e.g., SNe~2001el, 2002bo,
  see][]{2003ApJ...591.1110W,2007Sci...315..212W}.
Despite the fact that these two observations probe completely
different parts of the ejecta, and it is very difficult to link them
in a common conclusion, they might hint that a simple spherically
symmetric explosion cannot accurately describe SN~2003hv.  This is not
incompatible with our idea of clumping.  A complete treatment and
analysis of the spectropolarimetry will be given by \citet[][in
  prep.]{Maund2009}.

%%%%%%%%%%%%%%%%%%%%%%%%%%%%%  8. CONCLUSIONS   %%%%%%%%%%%%%%%%%%%%%%%%%%%%%%%%%%%%%%%%%

\section{Summary and conclusions}
\label{sec:conc}

We have presented observations of SN~2003hv that were obtained with a
multitude of instruments.  This study includes systematic multi-band
observations from early to very late phases, and the latest-ever
detection of a SN~Ia in the NIR.  It also features a comparison of our
nebular spectrum synthesis model with the widest (in wavelength
coverage) nebular spectrum of a SN~Ia.

SN~2003hv is a SN~Ia with the unusual value of $\Delta m_{15}(B) =
1.61 \pm 0.02$ mag, and it exhibits photometric and spectroscopic
properties that are consistent with its decline-rate parameter.

The main conclusions of the late-phase study are as follows:

\labelitemiii~
The individual light curves have decline rates similar to what has
been observed for other SNe~Ia in the past, confirming that there is
an evolution of the flux from the optical to the NIR wavelengths.

\labelitemiii~
At $+$700 days, a deceleration in the fading of the SN emission is
observed in the individual optical bands, particularly in the
$V$ band. Such tendencies have been seen in other SNe~Ia observed
past 600 days.

\labelitemiii~
 By comparing the radioactive energy input to the ejecta, as
  expressed from the \Nif\ masses derived from the UVOIR light curve
  at maximum light and in the tail, we find that the amount of energy
  probed within the UVOIR wavelengths is substantially less at
  late times than at maximum brightness. 

\labelitemiii~
A possible explanation for this could be that positrons escape the
ejecta, thus stealing energy away from it.  The UVOIR light curve,
however, follows very accurately the radioactive decay of \Cif\,
in the range $+$300--700 days, assuming complete and instantaneous positron
trapping. This slope is difficult to reconcile with positron escape.
Alternatively, the UVOIR light curve is not a good probe of the
  real bolometric light curve at late times, because the energy is
  still within the ejecta but a significant part of it is emitted at
  even longer wavelengths.

\labelitemiii~
A hypothesis proposed here to explain the SN 2003hv data is that an
IRC has occurred in the densest (i.e., innermost) part of the
ejecta. This idea can explain (i) the missing flux in the tail of the
UVOIR curve, (ii) the flat-topped NIR profiles, and (iii) the excess
flux observed in the MIR spectrum. However, such an IRC must have
taken place before $\sim$300--350 days, since this is when the slope
settles down, the flat-topped profiles start to emerge, and the 
{\em Spitzer} observation took place.

\labelitemiii~
The notion of an IRC occurring locally (and not simultaneously all
over the SN ejecta) is consistent with models that feature a clumped
(or inhomogeneous) distribution of the ejecta. The high-density
regions undergo the IRC first, since it is these that experience the
most efficient cooling.

\labelitemiii~
Deviations from spherical symmetry are suggested for SN~2003hv by the
blueshifts of the iron-family element lines from the optical to the
MIR and, independently, from early spectropolarimetric observations.

%%%%%%%%%%%%%%%%%%%%%%%%%%%%%  ACKNOWLEDGMENTS  %%%%%%%%%%%%%%%%%%%%%%%%%%%%%%%%%%%%

\begin{acknowledgements}
The Dark Cosmology Centre is funded by the Danish National Research
Foundation.  MS acknowledges support from the National Science
Foundation (NSF) under grant AST-0306969.  JS is a Royal Swedish
Academy of Sciences Research Fellow supported by a grant from the Knut
and Alice Wallenberg Foundation.  
AVF's group at UC Berkeley is supported by NSF grant AST-0607485 (UC
Berkeley), by the TABASGO Foundation, and by NASA/{\it HST} grants
GO-10272, AR-10952, and AR-11248 from STScI. KAIT and its ongoing
operation were made possible by donations from Sun Microsystems, Inc.,
the Hewlett-Packard Company, AutoScope Corporation, Lick Observatory,
the NSF, the University of California, the Sylvia \& Jim Katzman
Foundation, and the TABASGO Foundation.  AVF thanks the Aspen Center for Physics, where he participated in a
workshop on Wide-Fast-Deep Surveys while this paper was nearing completion.
CF acknowledges support from the Swedish National Space Board and the Swedish Research Council.  
MH is grateful for support provided by NASA through Hubble Fellowship grant
HST-HF-01139.01-A (awarded by the Space Telescope Science Institute
[STScI], which is operated for NASA by the Association of Universities
for Research in Astronomy, Inc., under contract NAS 5-26555), the
Carnegie Postdoctoral Fellowship, FONDECYT through grant 1060808, the
Millennium Center for Supernova Science through grant P06-045-F
(funded by ``Programa Bicentenario de Ciencia y Tecnolog\'ia de
CONICYT'' and ``Programa Iniciativa Cient\'ifica Milenio de
MIDEPLAN''), Centro de Astrof\'\i sica FONDAP 15010003, and Center of
Excellence in Astrophysics and Associated Technologies (PFB 06).
The SSO observations were obtained as part of the European Union's Human Potential Programme
``The Physics of Type Ia Supernovae'' under contract
HPRN-CT-2002-00303.  This research has made use of the Two Micron
All-Sky Survey (2MASS), and the NASA/IPAC Extragalactic Database
(NED), which is operated by the Jet Propulsion Laboratory, California
Institute of Technology, under contract with NASA.  We acknowledge
Bruno Leibundgut and Peter Lundqvist for early discussions on this
project. We are extremely grateful to Chris Gerardy for providing us
with the Spitzer data, to Kentaro Motohara for sending us the Subaru
NIR spectrum, and to Dietrich Baade and collaborators for providing an
advance look at their data from ESO program 71.D-0141(A).  GL wishes
to thank Ken Nomoto for useful discussions on SN~2003hv in Cefal\` u,
Sicily, and Keiichi Maeda for reading the manuscript and providing us
with comments.  We are also grateful to Sergio Gonzalez, Nidia
Morrell, and Miguel Roth for obtaining some observations of SN~2003hv.
Finally, we thank the referee, M.~Turatto, for providing  useful comments. 

\end{acknowledgements}

%%%%%%%%%%%%%%%%%%%%%%%%%%%%%  REFERENCES  %%%%% %%%%%%%%%%%%%%%%%%%%%%%%%%%%%%%%%%%

\bibliographystyle{aa}  %style aa.bst
\bibliography{Leloudas2003hv_final.bib}

{\small 
% if table 2
\longtab{4}{
\begin{longtable}[h]{ccclccccc}
\caption{\label{tab:phot} Optical photometry of SN~2003hv.$^a$}\\
\hline\hline
Date &MJD & Phase & Telescope   & $U$ & $B$  & $V$ & $R$ & $I$ \\
(UT) & (days) &  (days)      &     &  (mag)    &  (mag)   &   (mag)  & (mag)  &  (mag)  \\
\hline
\endfirsthead
\caption{continued.}\\
\hline\hline
Date &MJD & Phase & Telescope   & $U$ & $B$  & $V$ & $R$ & $I$ \\
(UT) & (days) &  (days)      &     &  (mag)    &  (mag)   &   (mag)  & (mag)  &  (mag)  \\
\hline
\endhead
\hline
\endfoot
2003 09 10   &  52892.40   &    1.2  &   CTIO 0.9~m  &    11.940(0.152) &   12.452(0.019) &	 12.483(0.015) &    12.459(0.015) &    12.759(0.015)	 \\
2003 09 10   &  52892.47   &    1.3  &   KAIT	      &       \nodata	  &   12.523(0.015) &	 12.565(0.015) &    12.489(0.015) &    12.827(0.026)	 \\
2003 09 11   &  52893.39   &    2.2  &   CTIO 0.9~m  &    12.125(0.102)  &   12.505(0.020) &	 12.510(0.015) &    12.477(0.015) &    12.779(0.015)	 \\
2003 09 11   &  52893.41   &    2.2  &   KAIT	      &       \nodata	  &   12.619(0.015) &	 12.616(0.015) &    12.509(0.015) &    12.870(0.019)	 \\
2003 09 12   &  52894.37   &    3.2  &   CTIO 0.9~m  &    12.176(0.058)  &   12.534(0.018) &	 12.546(0.015) &    12.514(0.015) &    12.844(0.015)	 \\
2003 09 13   &  52895.37   &    4.2  &   CTIO 0.9~m  &    12.131(0.063)  &   12.631(0.015) &	 12.573(0.015) &    12.586(0.015) &    12.949(0.015)	 \\
2003 09 14   &  52896.38   &    5.2  &   CTIO 0.9~m  &    12.371(0.055)  &   12.702(0.016) &	 12.594(0.015) &    12.626(0.015) &    12.971(0.015)	 \\
2003 09 14   &  52896.52   &    5.3  &   KAIT	      &       \nodata	  &   12.750(0.015) &	 12.676(0.015) &    12.693(0.015) &    13.040(0.020)	 \\
2003 09 15   &  52897.37   &    6.2  &   CTIO  0.9~m  &    12.422(0.052)  &   12.805(0.015) &	 12.681(0.015) &    12.742(0.015) &    13.079(0.015)	 \\
2003 09 15   &  52897.51   &    6.3  &   KAIT	      &       \nodata	  &   12.887(0.015) &	 12.748(0.015) &    12.814(0.015) &    13.180(0.020)	 \\
2003 09 16   &  52898.52   &    7.3  &   KAIT	      &       \nodata	  &   12.985(0.015) &	 12.804(0.015) &    12.911(0.015) &    13.263(0.023)	 \\
2003 09 17   &  52899.52   &    8.3  &   KAIT	      &       \nodata	  &   13.094(0.015) &	 12.858(0.015) &    13.010(0.015) &    13.288(0.024)	 \\
2003 09 19   &  52901.40   &   10.2  &   LCO  Swope  &    13.113(0.089)  &   13.319(0.021) &	 13.004(0.018) &       \nodata    &      \nodata	 \\
2003 09 19   &  52901.44   &   10.2  &   KAIT	      &       \nodata	  &   13.406(0.015) &	 13.044(0.015) &    13.120(0.015) &    13.330(0.019)	 \\
2003 09 20   &  52902.30   &   11.1  &   LCO  Swope  &    13.358(0.018)  &   13.410(0.015) &	 13.010(0.015) &    13.083(0.015) &    13.201(0.015)	 \\
2003 09 20   &  52902.48   &   11.3  &   KAIT	      &       \nodata	  &   13.520(0.015) &	 13.055(0.015) &    13.175(0.015) &    13.327(0.017)	 \\
2003 09 21   &  52903.46   &   12.3  &   KAIT	      &       \nodata	  &   13.664(0.015) &	 13.134(0.015) &    13.193(0.015) &    13.317(0.019)	 \\
2003 09 22   &  52904.49   &   13.3  &   KAIT	      &       \nodata	  &   13.789(0.015) &	 13.247(0.015) &    13.233(0.015) &    13.277(0.019)	 \\
2003 09 23   &  52905.26   &   14.1  &   CTIO 0.9~m  &    13.788(0.039)  &   13.825(0.015) &	 13.245(0.015) &    13.178(0.015) &    13.207(0.015)	 \\
2003 09 23   &  52905.30   &   14.1  &   LCO Swope  &    13.838(0.040)  &   13.851(0.015) &	 13.215(0.015) &    13.118(0.015) &    13.128(0.015)	 \\
2003 09 24   &  52906.40   &   15.2  &   LCO Swope  &    14.034(0.035)  &   14.007(0.015) &	 13.328(0.015) &    13.163(0.015) &    13.180(0.015)	 \\
2003 09 24   &  52906.48   &   15.3  &   KAIT	      &       \nodata	  &   14.106(0.015) &	 13.333(0.015) &    13.225(0.015) &    13.218(0.017)	 \\
2003 09 25   &  52907.30   &   16.1  &   LCO Swope  &    14.162(0.089)  &   14.157(0.016) &	 13.352(0.023) &    13.180(0.015) &    13.043(0.029)	 \\
2003 09 26   &  52908.30   &   17.1  &   LCO Swope  &    14.249(0.038)  &   14.298(0.015) &	 13.320(0.021) &    13.163(0.015) &    13.040(0.029)	 \\
2003 09 28   &  52910.47   &   19.3  &   KAIT	      &       \nodata	  &   14.621(0.015) &	 13.610(0.015) &    13.360(0.015) &    13.137(0.021)	 \\
2003 09 30   &  52912.47   &   21.3  &   KAIT	      &       \nodata	  &   14.808(0.015) &	 13.798(0.015) &    13.478(0.015) &    13.189(0.019)	 \\
2003 10 02   &  52914.40   &   23.2  &   LCO Swope  &    15.020(0.102)  &   14.992(0.022) &	 13.925(0.015) &    13.561(0.020) &    13.212(0.015)	 \\
2003 10 02   &  52914.46   &   23.3  &   KAIT	      &       \nodata	  &   15.022(0.015) &	 13.988(0.015) &    13.644(0.015) &    13.308(0.022)	 \\
2003 10 04   &  52916.46   &   25.3  &   KAIT	      &       \nodata	  &   15.158(0.015) &	 14.141(0.015) &    13.844(0.015) &    13.512(0.022)	 \\
2003 10 07   &  52919.42   &   28.2  &   KAIT	      &       \nodata	  &   15.340(0.015) &	 14.393(0.015) &    14.071(0.015) &    13.734(0.024)	 \\
2003 10 09   &  52921.40   &   30.2  &   KAIT	      &       \nodata	  &   15.448(0.015) &	 14.480(0.015) &    14.189(0.015) &    13.913(0.025)	 \\
2003 10 13   &  52925.43   &   34.2  &   KAIT	      &       \nodata	  &   15.640(0.021) &	 14.658(0.022) &    14.392(0.020) &    14.143(0.030)	 \\
2003 10 16   &  52928.42   &   37.2  &   KAIT	      &       \nodata	  &   15.646(0.015) &	 14.746(0.015) &    14.518(0.015) &    14.315(0.032)	 \\
2003 10 17   &  52929.40   &   38.2  &   LCO  Swope  &    15.511(0.045)  &   15.624(0.015) &	 14.781(0.015) &    14.523(0.015) &    14.325(0.015)	 \\
2003 10 18   &  52930.42   &   39.2  &   KAIT	      &       \nodata	  &   15.716(0.015) &	 14.806(0.015) &    14.593(0.015) &    14.428(0.025)	 \\
2003 10 20   &  52932.37   &   41.2  &   KAIT	      &       \nodata	  &   15.735(0.015) &	 14.871(0.015) &    14.671(0.015) &    14.566(0.022)	 \\
2003 10 22   &  52934.41   &   43.2  &   KAIT	      &       \nodata	  &   15.806(0.015) &	 14.916(0.015) &    14.752(0.015) &    14.629(0.018)	 \\
2003 10 24   &  52936.40   &   45.2  &   KAIT	      &       \nodata	  &   15.843(0.015) &	 15.018(0.015) &    14.840(0.015) &    14.735(0.025)	 \\
2003 10 27   &  52939.36   &   48.2  &   KAIT	      &       \nodata	  &   15.929(0.015) &	 15.094(0.015) &    14.926(0.015) &    14.879(0.025)	 \\
2003 10 30   &  52942.36   &   51.2  &   KAIT	      &       \nodata	  &   15.965(0.015) &	 15.199(0.015) &    15.049(0.015) &    15.043(0.031)	 \\
2003 10 30   &  52942.66   &   51.5  &   SSO 2.3~m   &    15.792(0.039)  &   15.952(0.016) &	 15.234(0.015) &    15.099(0.015) &    14.877(0.017)	 \\
2003 11 02   &  52945.38   &   54.2  &   KAIT	      &       \nodata	  &   16.053(0.015) &	 15.269(0.015) &    15.144(0.015) &    15.204(0.024)	 \\
2003 11 05   &  52948.32   &   57.1  &   KAIT	      &       \nodata	  &   16.142(0.059) &	 15.385(0.015) &    15.241(0.015) &    15.323(0.035)	 \\
2003 11 08   &  52951.30   &   60.1  &   LCO Swope  &    16.110(0.035)  &   16.086(0.015) &	 15.408(0.015) &    15.310(0.015) &    15.379(0.015)	 \\
2003 11 11   &  52954.36   &   63.2  &   KAIT	      &       \nodata	  &   16.168(0.026) &	 15.552(0.017) &    15.593(0.026) &       \nodata	 \\
2003 11 16   &  52959.33   &   68.1  &   KAIT	      &       \nodata	  &   16.246(0.017) &	 15.665(0.015) &    15.651(0.015) &    15.864(0.035)	 \\
2003 11 18   &  52961.28   &   70.1  &   CTIO 0.9~m  &       \nodata	  &   16.260(0.015) &	 15.667(0.015) &    15.644(0.015) &    15.855(0.015)	 \\
2003 11 19   &  52962.34   &   71.1  &   KAIT	      &       \nodata	  &   16.323(0.015) &	 15.776(0.015) &    15.744(0.015) &    15.928(0.032)	 \\
2003 11 20   &  52964.24   &   73.0  &   VLT Antu   &       \nodata	  &   16.277(0.042) &	 15.793(0.015) &    15.582(0.024) &    16.015(0.015)	 \\
2003 11 21   &  52964.32   &   73.1  &   CTIO 0.9~m  &       \nodata	  &   16.298(0.015) &	 15.815(0.015) &    15.764(0.015) &    15.991(0.015)	 \\
2003 11 22   &  52965.29   &   74.1  &   CTIO 0.9~m  &       \nodata	  &   16.347(0.015) &	 15.819(0.015) &    15.797(0.015) &    16.036(0.015)	 \\
2003 11 22   &  52965.31   &   74.1  &   KAIT	      &       \nodata	  &   16.338(0.015) &	 15.855(0.018) &    15.846(0.015) &	  \nodata	 \\
2003 11 27   &  52970.30   &   79.1  &   KAIT	      &       \nodata	  &   16.465(0.015) &	 15.982(0.016) &    16.047(0.015) &    16.285(0.047)	 \\
2003 11 28   &  52971.67   &   80.5  &   SSO 2.3~m   &    16.812(0.029)  &   16.372(0.015) &    15.935(0.015) &    15.921(0.017) &    15.981(0.024)     \\
2003 12 03   &  52976.28   &   85.1  &   KAIT	      &       \nodata	  &   16.576(0.033) &	 16.186(0.039) &    16.244(0.030) &    16.519(0.065)	 \\
2003 12 15   &  52988.30   &   97.1  &   LCO Swope  &    17.116(0.041)  &   16.694(0.015) &	 16.424(0.015) &    16.512(0.015) &    16.654(0.015)	 \\
2003 12 17   &  52990.26   &   99.1  &   KAIT	      &       \nodata	  &   16.836(0.024) &	 16.615(0.035) &    16.706(0.048) &    16.854(0.087)	 \\
2003 12 22   &  52995.10   &  103.9  &   LCO Swope  &    17.349(0.036)  &	 \nodata    &	 16.569(0.015) &    16.736(0.015) &    16.840(0.016)	 \\
2003 12 22   &  52995.24   &  104.0  &   KAIT	      &       \nodata	  &   16.869(0.019) &	 16.686(0.040) &    16.771(0.054) &    17.052(0.091)	 \\
2003 12 23   &  52996.10   &  104.9  &   LCO Swope  &       \nodata	  &   16.890(0.015) &	    \nodata    &       \nodata    &	  \nodata	 \\
2003 12 27   &  53000.10   &  108.9  &   LCO Swope  &    17.549(0.032)  &   16.918(0.015) &	 16.718(0.015) &    16.889(0.021) &    17.012(0.017)	 \\
2003 12 28   &  53001.21   &  110.0  &   KAIT	      &       \nodata	  &   17.010(0.023) &	 16.817(0.033) &    17.085(0.027) &    17.414(0.085)	 \\
2003 12 28   &  53001.57   &  110.4  &   SSO 2.3~m   &    17.588(0.015)  &   16.859(0.018) &    16.786(0.015) &    16.936(0.015) &    17.119(0.016)     \\
2004 01 06   &  53010.20   &  119.0  &   KAIT	      &       \nodata	  &   17.120(0.052) &	 17.107(0.047) &    17.336(0.048) &    17.316(0.100)	 \\
2004 01 09   &  53013.19   &  122.0  &   KAIT	      &       \nodata	  &   17.254(0.128) &	 17.208(0.090) &    17.329(0.108) &	  \nodata	 \\
2004 01 13   &  53017.18   &  126.0  &   KAIT	      &       \nodata	  &   17.241(0.019) &	 17.233(0.024) &    17.476(0.037) &    17.576(0.081)	 \\
2004 01 16   &  53020.16   &  129.0  &   KAIT	      &       \nodata	  &   17.299(0.026) &	 17.242(0.026) &    17.567(0.020) &    17.694(0.111)	 \\
2004 01 19   &  53023.13   &  131.9  &   KAIT	      &       \nodata	  &   17.349(0.028) &	 17.347(0.026) &    17.625(0.041) &    17.706(0.072)	 \\
2004 01 22   &  53026.15   &  135.0  &   KAIT	      &       \nodata	  &   17.456(0.025) &	 17.379(0.033) &    17.638(0.037) &    17.769(0.092)	 \\
2004 01 29   &  53033.49   &  142.3  &   SSO 2.3~m   &    18.440(0.057)  &   17.518(0.043) &	 17.373(0.024) &    17.581(0.032) &    17.686(0.101)	 \\
2004 03 29   &  53093.38   &  202.2  &   SSO 2.3~m   &    19.843(0.330)  &   18.634(0.055) &	 18.383(0.050) &    18.691(0.100) &    18.276(0.147)	 \\
2004 07 10   & 53196.55    &  305.4   &  {\it HST} ACS$^b$    &       \nodata	  &   \nodata   		&	 19.896(0.021) 	&       \nodata     	&    20.052(0.017) \\
2004 07 12  & 53198.82    &  307.6    &  {\it HST} ACS    &       \nodata	  &    20.058(0.017)  	&	  \nodata           	&    21.210(0.029)    	&     \nodata	    \\
2004 07 16  & 53202.75    &  311.6    &  {\it HST} ACS    &       \nodata	  &    20.079(0.016)	&	  \nodata            &    21.219(0.024)	&     \nodata	 \\
2004 08 12   & 53229.30    &  338.1  &   VLT Kueyen  &    21.648(0.039)  &   20.508(0.028) &	 20.543(0.033) &    21.224(0.042) &    20.587(0.032)	 \\
2004 08 13   & 53230.39    &  339.2  &   VLT Kueyen  &       \nodata	  &	 \nodata    &	    \nodata    &    21.263(0.034) &    20.592(0.032)	 \\
2004 11 15  & 53324.71    &  433.5    &  {\it HST} ACS    &       \nodata	  &   \nodata   		&	 21.712(0.029) 	&       \nodata     	&    21.417(0.031) \\
2005 02 07   & 53408.10    &  516.9  &   VLT Antu    &       \nodata	  &	 \nodata    &	 23.326(0.100) &    23.111(0.102) &    22.554(0.122)	 \\ 
2005 03 04   & 53433.02    &  541.8  &   VLT Antu    &    $>$22.95       &	 \nodata    &	    \nodata    &       \nodata    &	  \nodata	 \\ 
2005 08 08   & 53590.34    &  699.1  &   VLT Kueyen  &       \nodata	  &   25.198(0.137) &	    \nodata    &       \nodata    &    24.366(0.287)$^c$	 \\
2005 08 09   & 53591.36    &  700.2  &   VLT Kueyen  &       \nodata	  &	 \nodata    &	    \nodata    &    25.192(0.388) &	  \nodata	 \\
2005 08 10   & 53592.39    &  701.2  &   VLT Kueyen  &       \nodata	  &	 \nodata    &	 24.423(0.138) &       \nodata    &	  \nodata	 \\
\end{longtable}
\begin{tabular}{lll}
$^a$The quoted values do not include S-corrections. Numbers in parentheses 
are uncertainties. All formal errors below 0.015 have been \\
rounded up to this value.\\
 $^b$Vega magnitudes in the {\it HST} filter system ($F435W$,
  $F555W$, $F625W$, and $F814W$).\\
$^c$In the last VLT epoch, two $R$ images and two $I$ images obtained at 
epochs differing by 5 and 1 days (respectively) have been combined \\
to  increase the signal-to-noise ratio.
\end{tabular}
}	% End \longtab
}	% End \small

%%%%%%%%%%%%%%%%%%%%%%%%%%%%%%%%%%%%%% APPENDIX %%%%%%%%%%%%%%%%%%%%%%%%%%%%%%%%%

\clearpage
\begin{appendix}
\section{Observation logs}

\begin{table}[h]
\caption{VLT late-time optical observations of SN~2003hv.}
\label{tab:VLTopticallog}
\centering
\begin{tabular}{cccccccc}
\hline\hline
Date & MJD &Phase & Filter & Exposure & Airmass & Seeing & Instrument  \\
(UT) & (days)      &  (days)    &   & (s)      &         & (arcsec) & \\
\hline
2004 08 12   &   53229.33     &   338.1    & $U$    & 3$\times$1000& 1.17 & 1.02 & FORS1 \\
2004 08 12   &   53229.37     &   338.2    & $B$    & 3$\times$420 & 1.06 & 0.85 & FORS1  \\
2004 08 12   &   53229.39     &   338.2    & $V$    & 3$\times$300 & 1.03 & 0.74 & FORS1  \\
2004 08 12   &   53229.40     &   338.2    & $R$    & 2$\times$420 & 1.01 & 0.75 & FORS1  \\
2004 08 12   &   53229.41     &   338.2    & $I$    & 1$\times$500 & 1.01 & 0.70 & FORS1  \\
2004 08 13   &   53230.36     &   339.2    & $I$    & 4$\times$500 & 1.08 & 0.78 & FORS1  \\
2004 08 13   &   53230.39     &   339.2    & $R$    & 2$\times$420 & 1.03 & 0.76 & FORS1 \\
2005 02 07   &   53408.03     &   516.8    & $I$    &12$\times$240 & 1.13 & 1.13 & FORS2  \\
2005 02 07   &   53408.08     &   516.9    & $R^a$  & 3$\times$600 & 1.37 & 0.93 & FORS2  \\
2005 02 07   &   53408.09     &   516.9    & $V$    & 3$\times$400 & 1.49 & 0.98 & FORS2  \\
2005 03 04   &   53433.02     &   541.8    & $U^a$  &3$\times$1140 & 1.48 & 1.24 & FORS2 \\
2005 08 04   &   53586.40     &   695.2    & $R$    & 2$\times$900 & 1.04 & 0.70 & FORS1\\
2005 08 07   &   53589.41     &   698.2    & $I$    & 2$\times$900 & 1.02 & 0.75 & FORS1\\
2005 08 08   &   53590.31     &   699.1    & $I$    & 4$\times$900 & 1.39 & 0.82 & FORS1\\ 
2005 08 08   &   53590.34     &   699.1    & $B$    & 5$\times$540 & 1.19 & 0.82 & FORS1\\
2005 08 09   &   53591.36     &   700.2    & $R$    & 3$\times$900 & 1.12 & 0.62 & FORS1\\
2005 08 10   &   53592.39     &   701.2    & $V$    & 3$\times$480 & 1.03 & 0.71 & FORS1\\
\hline
\end{tabular} \\
\begin{tabular}{lll}
$^a$The $U$ and $R$ filters on FORS2 are slightly different than the 
ones on FORS1.\\
\end{tabular}
\end{table}

%\clearpage
\begin{table}[h]
\caption{VLT late-time NIR observations of SN~2003hv.}
\label{tab:VLTIRLOG}
\centering
\begin{tabular}{ccccccc}
\hline\hline
Date & MJD & Phase & Filter & Exposure$^a$ & Airmass & Seeing  \\
(UT) &(days)     &  (days) &        & (s)      &         & (arcsec) \\
\hline
2004 08 18  & 53235.37    &  344.2  & $J_{s}$	  &   30$\times$4$\times$24 & 1.04 & 0.55\\  
2004 08 29  & 53246.25    &  355.1  & $K_{s}$	  &   10$\times$6$\times$30 & 1.37 & 0.47 \\
2004 08 29  & 53246.29    &  355.1  & $H$	  &   10$\times$6$\times$59 & 1.17 & 0.75\\  
2004 08 29  & 53246.32    &  355.1  & $K_{s}$	  &   10$\times$6$\times$30 & 1.07 & 0.57 \\    
2004 08 29  & 53246.39    &  355.2  & $H$	  &   10$\times$6$\times$30 & 1.00 & 0.75 \\ 
2004 08 30  & 53247.26    &  356.1  & $J_{s}$	  &   30$\times$4$\times$23 & 1.29 & 0.77 \\
2004 08 30  & 53247.30    &  356.1  & $K_{s}$	  &   10$\times$6$\times$30 & 1.13 &  0.50  \\
2005 01 24  & 53394.02    &  502.8  & $K_{s}$	  &   10$\times$6$\times$60 & 1.03 &  0.47\\
2005 01 24  & 53394.09    &  502.9  & $H$	  &   10$\times$6$\times$30 & 1.24 &  0.74 \\
2005 02 24  & 53425.99    &  534.8  & $K_{s}$	  &   10$\times$6$\times$30 & 1.18 & 0.62 \\
2005 02 25$^b$  & 53426.99    &  535.8  & $J_{s}$	  &   30$\times$4$\times$46 & 1.19 & 0.83 \\  
2005 02 27  & 53428.99    &  537.8  & $H$	  &   10$\times$6$\times$30 & 1.22 & 0.55 \\
2005 02 28  & 53429.99    &  538.8  & $J_{s}$	  &   30$\times$4$\times$23 & 1.22 & 0.56\\
2005 10 13  & 53656.18    &  765.0  & $J_{s}$	  &   30$\times$4$\times$46 & 1.11 & 0.70 \\
2005 10 13  & 53656.35    &  765.2  & $J_{s}$	  &   30$\times$4$\times$23 & 1.14 & 0.56 \\    
2005 10 14  & 53657.21    &  766.0  & $J_{s}$	  &   30$\times$4$\times$46 & 1.04 & 0.54\\ 
2005 10 14  & 53657.32    &  766.1  & $H$	  &   10$\times$6$\times$60 & 1.05 & 0.72 \\ 
2005 10 15  & 53658.31    &  767.1  & $H$	  &   10$\times$6$\times$30 & 1.04 & 0.40\\  
  \hline
\end{tabular} 
\begin{tabular}{lll}
$^a$Detector integration time (DIT) $\times$ number of DITs per 
exposure $\times$ number of exposures. \\
$^b$Image not included in the final photometry as it is a poor image and 
gives bad results. \\
\end{tabular}
\end{table}

%\clearpage
\begin{table*}[hp]
\caption{{\it HST} observations of SN~2003hv.$^a$}
\label{tab:HSTlog}
\centering
\begin{tabular}{ccccccc}
\hline\hline
Date & MJD &Phase & Instrument & Filter & Exposure   \\
(UT) & (days)      & (days)   &    &  & (s)       \\
\hline
2004 07 10  & 53196.55    &  305.4   &ACS   & $F555W$  & 480  \\
2004 07 10  & 53196.56    &  305.4    &ACS  & $F814W$  & 720  \\
2004 07 12  & 53198.82    &  307.6   &ACS   & $F435W$  & 840  \\
2004 07 12  & 53198.83    &  307.6    &ACS  & $F625W$  & 360  \\
2004 07 16  & 53202.75    &  311.6    &ACS  & $F435W$  & 840   \\
2004 07 16  & 53202.76    &  311.6    &ACS  & $F625W$  & 360  \\
2004 11 15  & 53324.71    &  433.5    &ACS  & $F555W$  & 480  \\
2004 11 15  & 53324.72    &  433.5    &ACS  & $F814W$  & 720   \\
2005 11 03  & 53677.31    &  786.1     &NICMOS  & $F160W$  &  $16 \times 640$   \\
\hline
\end{tabular} \\
\begin{tabular}{lll}
$^a$Observations not conducted due to guide-star failure are not included.\\
\end{tabular}
\end{table*}

\end{appendix}

\end{document}